\documentclass[journal]{IEEEtran}
\usepackage{lipsum}
\usepackage{longtable}
\usepackage{ifpdf}
\usepackage{cite}
\usepackage{algorithmic}
\usepackage{algorithm}

\ifCLASSINFOpdf
   \usepackage[pdftex]{graphicx}
 
\else
 \usepackage[dvips]{graphicx}
\fi
\usepackage{epsfig}
\usepackage[cmex10]{amsmath}
\usepackage{epstopdf}
\usepackage{array}
\usepackage{flushend}
\usepackage{balance}
\usepackage[top = 0.75 in, bottom = 0.5450in, left = 0.7in, right = 0.721181in]{geometry}
\usepackage{eqparbox}

\usepackage{fixltx2e}
\usepackage{color}
\usepackage{dblfloatfix}
\usepackage{url}
\ifCLASSOPTIONcompsoc
    \usepackage[caption=false, font=normalsize, labelfont=sf, textfont=sf]{subfig}
\else
\usepackage[caption=false, font=footnotesize]{subfig}
\fi

\usepackage{tabularx}
\usepackage{amssymb}
\begin{document}

\title{On Optimizing VLC Networks for Downlink Multi-User Transmission: A Survey}

\author{Mohanad~Obeed,~\IEEEmembership{Student Member,~IEEE,}
        Anas~M.~Salhab,~\IEEEmembership{Senior Member,~IEEE,}
        Mohamed-Slim~Alouini,~\IEEEmembership{Fellow,~IEEE}
        and~Salam~A.~Zummo,~\IEEEmembership{Senior Member,~IEEE,}}

\maketitle

\begin{abstract}
The evolving explosion in high data rate services and applications  will soon require  the use of untapped, abundant unregulated spectrum of the visible light  for communications to adequately meet the demands of the fifth-generation (5G) mobile technologies. Radio-frequency (RF) networks are proving to be scarce to cover the escalation in data rate services. Visible light communication (VLC) has emerged as  a great potential solution, either in replacement of, or  complement to,  existing RF networks,  to support the projected traffic demands. Despite of the prolific advantages of VLC networks, VLC faces many challenges that must be resolved  in the near future to achieve a full standardization and to be integrated  to future wireless systems. Here, we review the new, emerging research   in the field of VLC networks  and lay out the challenges,  technological solutions, and future work predictions. Specifically, we  first review the VLC channel capacity derivation, discuss the performance metrics and the associated variables;  the optimization of  VLC networks are also discussed,  including  resources and power allocation techniques, user-to-access point (AP) association and APs-to-clustered-users-association, APs coordination techniques, non-orthogonal multiple access (NOMA) VLC networks, simultaneous energy harvesting and information transmission using the visible light, and the security issue in VLC networks.  Finally,  we propose several open research problems to optimize the various VLC networks by maximizing either the sum rate, fairness, energy efficiency, secrecy rate, or harvested energy.   
\end{abstract}
\IEEEpeerreviewmaketitle

\section{Introduction}

With the dramatic increase in  total data traffic (approximately 7.24 exabyte-per-month in 2016,  predicted  to be 48.95 exabyte-per-month in 2021 \cite{Cisco}), there is an urgent need to develop  a fifth-generation (5G) of networks with a higher system-level spectral efficiency that will  offer  higher data rates, massive device connectivity, higher energy efficiency (EE), lower traffic fees, a more robust security, and ultra-low latency \cite{andrews2014will,shafi20175g, pirinen2014brief}. With the advent of the internet-of-things (IoT) era, the amount of the connected devices to the internet is increasing dramatically \cite{al2015internet, palattella2016internet}, resulting in a significant increase in data traffic that, and hence, crowded traditional radio-frequency (RF) or wireless-fidelity (WiFi)  networks \cite{tsonev2015towards}.  Small cells or network densification have been proposed as a solution for 5G technologies \cite{bhushan2014network, ge20165g} in order to increase the system capacity and coverage, reduce the power consumption of mobile devices, and enhance the networks' EE.  The continuity of dramatic growing in data traffic demand has motivated researchers to explore new spectrum, new techniques, and new network architectures to meet these demands. Visible light communication (VLC) has been introduced as a promising solution for 5G and beyond. The motivation behind emerging the VLC technology is the great invention of the energy-efficient light emitting diode (LED) \cite{tanaka2000wireless}. White LEDs outperform the other light sources with their modulation performance, high electrical-to-optical conversion efficiency, long life span, small size and light weight, low cost, and operational speed \cite{kishi2014high, siddique2013joint, wir}. LED lamps consume approximately 20\%  of the power consumed by  fluorescent bulbs and approximately 0.5\% of the power consumed by traditional light sources \cite{kavehrad2010sustainable}.

Visible light communication uses a portion of the electromagnetic spectrum that is entirely untapped, free, safe, and provides a high potential bandwidth for wireless data transmission with rejecting the present RF interference \cite{elgala2011indoor}.  Hence, VLC is a communication technology that uses LEDs as  transmitters to emit both the light and information signals  to the users. We should note that the power of the information signal must meet the illumination requirements, as well as being within the range of the LED's physical limits \cite{din2014energy}. However, the non-linearity of LEDs in electrical-to-optical transfer can be efficiently avoided using pre-distortion mechanisms \cite{dimitrov2013information}. The VLC receiver contains a photo-detector (PD) component that has the ability to convert the received light intensity to a current signal. Data are transmitted using an intensity modulation (IM) technique at the transmitter,  and received using a direct detection (DD) technique at the receiver (IM/DD) \cite{dimitrov2015principles}. This means that the modulating signal must be real non-negative, and the existing modulation techniques in the RF networks adjusted to fit this property.  Compared to RF networks,  VLC networks provide higher data rates, larger EEs, lower battery consumption, and smaller latency. In addition, VLC can be safely used in sensitive environments such as chemical plants, aircraft, and hospitals \cite{miyakoshi2013cellular}. Because of the small coverage of the transmitters in VLC systems, an exhaustive reuse of frequency can be implemented.  VLC is also power-saving since the consumed power for communication is already used for illumination and may also be used for energy harvesting. Because the light can be blocked by opaque objects,  VLC functions properly only in line-of-sight (LoS) communications, which own a robust security since the unauthorized users who are out of sight cannot receive an information signal of good quality. 

Despite  all the aforementioned VLC advantages, VLC faces many technical challenges that must be resolved in the near future to achieve its full standardization  and  integration with future wireless systems. Among the most important challenges to be overcome are relatively small bandwidth of LEDs, deriving the exact channel capacity, channel estimation and shadowing effects,  backhauling VLC traffic into a large-scale networks, the rapid decrease in light intensity with distance, and the noise or interference that may be generated by nearby lighting systems. One  common solution to partially overcome these challenges is optimizing the parameters of VLC networks. Another common solution is to supplement the VLC network by RF networks. 

Numerous studies have investigated the potential applications of VLC to  outdoor communications; yet, VLC  is  better suited for indoor communications. According to various published statistics,  users of wireless information spend 80\% and 20\% of their time in indoor and outdoor environments, respectively \cite{feng2016applying}. In general, this paper reviews the optimization techniques studied in the literature to improve VLC systems' performance with focusing on target research directions. 

\begin{table}[!htbp]
\centering
\label{table_LoA}
\caption{ List of abbreviations}
\begin{tabular}{|p{.15\textwidth} | p{.3\textwidth} |}
\hline
4G& Fourth generation\\
\hline
5G & Fifth generation\\
\hline
AC & Alternative current\\
\hline
ACO-OFDM & Asymmetrically clipped optical OFDM\\
\hline
AP & Access point\\
\hline
APA & Access point assignment\\
\hline
BER & Bit error rate\\
\hline

CoMP & Coordinated multi-point \\
\hline
CSI & Channel-state-information\\
\hline
CSK & Color shift keying\\
\hline
DD & Direct detection\\
\hline
DC & Direct current\\
\hline
DCO-OFDM & Direct current optical OFDM\\
\hline
EE & Energy efficiency\\
\hline
EGT & Evolutionary game theory\\
\hline
EH & Energy harvesting\\
\hline

FoV & Field-of-view\\
\hline
FFR & Fractional frequency reuse\\
\hline
FR & Frequency reuse\\
\hline

GEE & Global energy efficiency\\
\hline
IM & Intensity modulation\\
\hline
IoT & Internet of things \\
\hline
LB & Load balancing\\
\hline
LED & Light emitting diode\\
\hline

LiFi & Light fidelity \\
\hline
LoS & Line of sight \\
\hline
MIMO & Multiple input multiple output \\
\hline
MINLP & Mixed-integer nonlinear programming\\
\hline
MISO & Multiple input single output \\
\hline
MRC & Maximum ration combining\\
\hline

MPPM & Multipulse pulse position modulation\\
\hline
NGDP & Normalized gain difference power allocation\\
\hline

NOMA & Non-orthogonal multiple access \\
\hline

OFDM & Orthogonal frequency division multiplexing\\
\hline
OFDMA & Orthogonal division multiple access\\
\hline
OMA & Orthogonal multiple access\\
\hline
OOK & On-off keying\\
\hline
OPPM & Overlapping pulse width modulation\\
\hline
OWC & Optical wireless communication\\
\hline
PD & Photo-detector\\
\hline
PD-NOMA & Power domain NOMA\\
\hline
PDMA & Pattern division multiple access\\
\hline

PIN & Positive-intrinsic-negative \\
\hline
PLC & Power line communication\\
\hline
PLS & Physical layer security\\
\hline
PPM & Pulse position modulation\\
\hline
PWM & pulse width modulation \\
\hline
QoS & Quality-of-service\\
\hline
RA & Resource allocation \\
\hline
 RF & Radio frequency  \\
 \hline 
 RGB & Red, green, and blue\\
 \hline
 RLL & Run length limited \\
 \hline
 SCMA & Sparse code multiple access\\
 \hline
 SDMA & Space-division-multiple-access\\
 \hline
 SIC & Successive interference cancelation \\
 \hline
 SINR & signal-to-noise and interference ratio\\
 \hline

 SLIPT & Simultaneous lightwave for information and power transfer\\
 \hline
 SNR & Signal-to-noise-ratio\\
 \hline
 SWIPT & Simultaneous wireless for information and power transfer\\
 \hline
 TDMA & Time division multiple access\\
 \hline

 VLC & Visible light communication \\
 \hline
 
 WiFi & Wireless-fidelity\\

  \hline
\end{tabular} 

 \end{table}

 \subsection{Related Work}
 Several review articles have been written in the past on the topic of the VLC technology \cite{kumar2010led,sevincer2013lightnets,wu2014visible, karunatilaka2015led,surveyvlc,li2018optimization,chowdhury2018comparative,qiu2016channel, sindhubala2016survey,zhuang2018survey,cuailean2017current}, but none of them addressed how the new emerging technologies in RF networks could be mapped and applied in VLC networks such as the non-orthogonal multiple access (NOMA), energy harvesting (EH), simultaneous wireless information and power transfer (SWIPT), space division multiple access (SDMA), and physical layer security (PLS).  
Specifically,  Kumar \emph{et al.}  reviewed LED-based VLC systems and applications in their early stage development \cite{kumar2010led}.  In \cite{sevincer2013lightnets}, authors focused on the dual function of LEDs (used in  smart lighting and VLC), and explored their potential for integration  by introducing a new concept: LIGHTNETs (LIGHTing and NETworking) that performs both functions simultaneously.  Authors of \cite{wu2014visible} highlighted the benefits and disadvantages of VLC networks, compared with RF networks. The benefits of LEDs for illumination and communications, modulation schemes, dimming techniques, and the methods used for improving the performance of VLC were reviewed in \cite{karunatilaka2015led}, while in \cite{surveyvlc}, authors focused on the VLC link level transmission and shed some light on medium access techniques and visible light sensing.  A more recent study by Li \emph{et al.} reviewed system-level VLC networks, with a focus on user-centric network design, and compared it with the network-centric design with emphasizing on the interference reduction techniques  \cite{li2018optimization}. 
In \cite{chowdhury2018comparative}, authors explored the differences among optical wireless communications (OWC) technologies such as infrared communications, VLC, light-fidelity (LiFi), free space optical (FSO) communications, etc.  

Some review articles focused on specific aspects of VLC such as VLC channel modeling methods \cite{qiu2016channel}, noise optical sources and noise mitigation mechanisms \cite{sindhubala2016survey},  VLC-based positioning techniques for indoor and outdoor applications \cite{zhuang2018survey}, and the pertinent issues associated with the outdoor usage of VLC in vehicular communication \cite{cuailean2017current}. They generally identified emerging challenges and proposed future research directions.

This paper explores all the optimization techniques, previously reported in the literature, that  aim to improve the VLC network performance. Emphasis is placed on  how the new technologies, emerged in RF networks, mapped or used in VLC networks such as NOMA, SWIPT, cooperative transmission, SDMA, and physical layer security.  

Specifically,

\begin{itemize}
\item This paper provides, in Section II, an overview of VLC technology, defines  and discusses the objectives and constraints that must be taken into account when optimizing VLC networks.  Special emphasis is placed on channel capacity derivations,  and the unique properties of VLC. We also discuss the variables, parameters, and constraints having an impact on the performance of VLC networks.  
\item All optimization techniques are reviewed in Section III,  including power and resource allocation, users-to-APs association, cell formation, and AP cooperation used for mitigating the disadvantages of VLC networks  to improve  performance. This important topic  was previously investigated by Li \emph{et al.}\cite{li2018optimization}. However, their study was focused  on the difference between  user-centric and network-centric cell formations, and the interference reduction techniques, whereas  in this paper, we place our attention on the techniques, used in RF/VLC and in VLC standalone networks, that are aimed at  alleviating the limitations in VLC networks. In other words, we show how to formulate optimization problems, what are the techniques used for solving these optimization problems, how the different objectives, limitations, constraints are evaluated, and how added RF APs can remove stand-alone VLC network limitations. 

\item  By reviewing all the work conducted on NOMA-VLC systems in Section IV, we show how such systems are  different from NOMA-RF architectures. 

\item In Section V, we survey the various energy harvesting techniques used in VLC networks  and show how this added function (energy harvesting) affects the illumination and communication functions that are implemented simultaneously, using  LEDs.    

\item The topic of  physical layer security in VLC networks is also reviewed in Section VI, including the different techniques proposed to improve the secure VLC communications.  

\item In Section VII, we outline some remaining challenges and open research areas in NOMA-VLC networks, energy harvesting in VLC systems, and securing VLC networks. We present several ideas, which have not been previously investigated or proposed, to improve the performance of the different types of VLC networks. 
\end{itemize}

\begin{figure}[h]
\centering

\includegraphics[clip, trim=9cm 1cm 0 0cm,width=14cm]{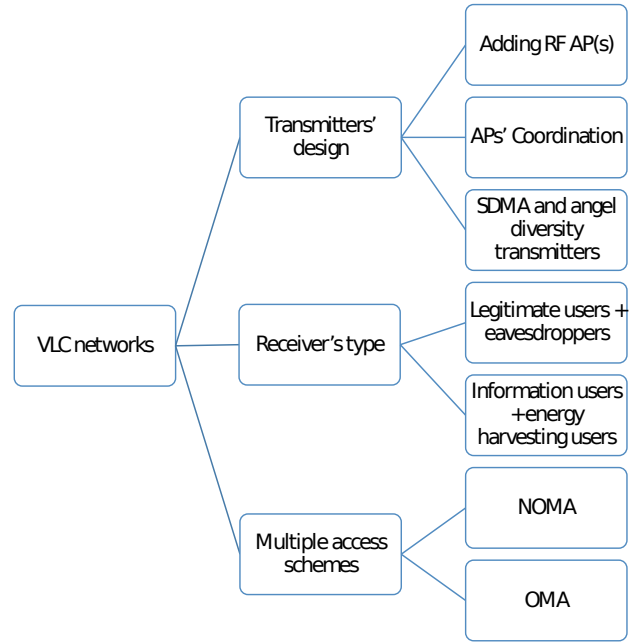}
\caption{Different types of VLC networks}
\label{chart}
\end{figure}

With this article, our goal is to present a clear, comprehensive picture of what has been accomplished so far, in the field of VLC networks, and to present future research directions.  
A list of abbreviation used in this paper is presented in Table \ref{table_LoA}, and the different types of VLC networks that considered in this paper are shown in Fig. \ref{chart}.

\section{Fundamentals of VLC indoor systems}

Because of its unique properties, a VLC channel is different from a RF or any other communication technology; its optical signal is modulated via the intensity of the signal, without  carrying any information in phase or in frequency; the transmitted signal is positive and real,  the optical power is proportional to the mean of the input power signal (not to the mean square of the signal amplitude); the transmitted peak power    is constrained by the LED’s dynamic range and the illumination requirements.

\subsection{VLC Elements}

Every  communication system must consist of a transmitter, channel, and receiver. Here, we discuss the characteristics of a VLC's transmitter and receiver.

 
\subsubsection{VLC Transmitter}
The  LED lamp  is the most appropriate transmitter used for both illumination and communication purposes (see Introduction for details). Each lamp usually consists of one or multiple LEDs driven by a circuit that controls the intensity of the brightness, using  the  the 'flowing-in' current. The function of the driver circuit is to transmit the data by modifying the flowing-in current, which, in turn,  modifies the light intensity. The flowing-in current must be within the LED’s dynamic range in order for the output (light intensity)  to be linearly proportional to the input current.  Because it shows the objects as they are without changing their real colors, the white color is commonly used for illumination and communication. 
Two common schemes are generally used in design white LEDs. One uses  a blue LED with a yellow Phosphor layer \cite{khalid20121}, the other uses a combination of three LEDs (red, green, and blue) \cite{cossu20123}. 



Because of its low cost and simplicity of implementation, the first
type of LEDs (the blue LED with a yellow phosphor layer) is more popular than the RGB type
for designing white LEDs. However, it has a limited bandwidth, compared to RGB, because of the slow absorption and emission of the coating phosphor layer. Khalid \emph{et al.} \cite{khalid20121} showed that a 1 Gbps data rate could be achieved, using this type of LEDs. 
 The RGB technique is better for communication as it uses the color shift keying (CSK) modulation technique that modulates the signal, using the three different LEDs. By doing so, data rates of 3.4 Gbps data can be achieved \cite{cossu20123}. 

One important issue that should be considered, when designing the VLC, is the illumination requirements, which is the main purpose of the LED. In other words, the illumination range that is required should not be violated by the VLC system. This means that the performance of the VLC system is related to the illumination design requirements (more details are given in Section \ref{RIC}).  
\subsubsection{VLC Receiver}

The PD is a diode device  sensitive to the light intensity that can convert the received light to a current modulated by the intensity of the light received. The PDs that are  commercially available can easily take samples of the received visible light at a  rate of tens of MHz \cite{surveyvlc}.

There are three types of devices that can be used as   VLC receivers of the optical signal coming from the LED\ transmitter: 1) photo-detector (e.g. positive-intrinsic-negative (PIN) and avalanche PD), 2) an imaging or camera sensor,  3) and a solar panel.   

 One of the advantages of a camera sensor is its availability on most mobile devices
such as smart-phones used to capture videos and images. The main advantage of a solar panel is that it can directly convert the received light to an electrical signal without the need for an external power supply \cite{wang2015design}.

\subsection{Channel Model}
\label{CM}
\begin{figure}[t]
\centering
\includegraphics[width=0.95\linewidth]{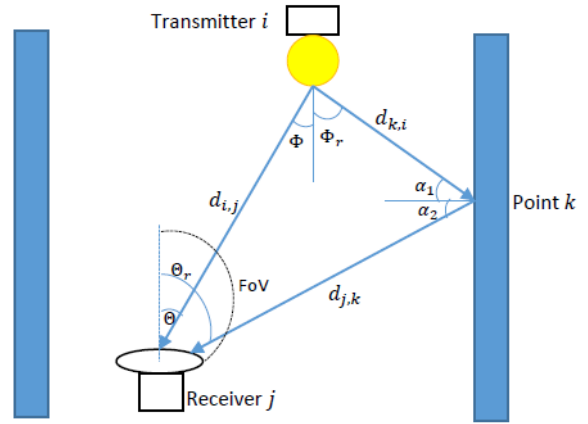}
\caption{Channel model, including the LoS link and the first reflected link}
\label{chan_fig}
\end{figure}

The receiver receives the LoS optical signal ans many copies of non-LoS, coming from reflections. According to \cite{komine2004fundamental}, the optical power received from signals reflected more than once is negligible. Fig \ref{chan_fig} shows a channel model of VLC links, containing the LoS and first reflected links.  
The LoS VLC link between the AP $i$ and the user $j$ can be expressed as follows \cite{wir, gfeller1979wireless}:

\begin{equation}
\label{vlcch}
h_{j,i}=\frac{A_{p}(m+1)}{2\pi d_{j,i}^2} \cos^m(\phi)g_{of}\cos(\theta)f(\theta),
\end{equation}%
where  $A_p$ is the physical area of the receiver PD, $m$ is the Lambertian index given by  $m=\frac{-1}{\log_2(\cos(\theta_{1/2})}$, with $\theta_{1/2}$  the half-intensity radiation angle, $d_{j,i}$  the distance between  AP $i$ and user $j$,  $g_{of}$   the gain of the optical filter, $\phi$  the  angle of
irradiance at the AP, $\theta$   the angle of incidence at the PD, and $f(\theta)$  the optical concentrator gain is given by
\begin{equation}
f(\theta)=\begin{cases}\frac{n^2}{\sin^2(\Theta)}, & 0\leq\theta\leq\Theta; \\
0, & \theta>\Theta,
\end{cases}
\end{equation}
where $n$ is the refractive index and $\Theta$ is the semi-angle
of the field-of-view (FoV) of PD. Komine and Nakagawa  \cite{komine2004fundamental} showed that the DC  attenuation of the channel, from the first reflected link is given by 

\begin{equation}
dh_1= \frac{(m+1)A_{p}}{2\pi d_{k,i}^2  d_{j,k}^2} \rho dA_s \cos^m(\phi_r)cos(\alpha_1)cos(\alpha_2)g_{of}f(\theta_r)\cos(\theta_r),
\end{equation} 
where $\alpha_r$ and $\theta_r$ are the angels of the irradiance and incidence of the first reflection link, respectively, $d_{k,i}^2$ and $d_{j,k}^2$ are the distance from the AP $i$ to the reflecting point $k$ and the distance from the reflecting point $k$ to the user $j$, respectively, $\rho$ and  $dA_s$ are the reflection factor and the reflective area, respectively, $\alpha_1$ and $\alpha_2$ are the  irradiance angles with respect to the reflected point and with respect to the receiver, respectively. 

\subsection{VLC Modulation Schemes}

As mentioned previously, data cannot be transmitted by encoding the phase or frequency, and the modulation in VLC is implemented by varying the light intensity of the LED. On the other hand, the demodulation can be implemented by  direct detection at the PD. Various IM/DD-based modulation techniques have been proposed and published in the literature. On-off keying (OOK) was proposed for VLC,  as a simple modulation scheme, where data are represented by two levels of light intensity \cite{rajagopal2012ieee, le2009100}.  In order to obtain higher data rates, in comparison with what OOK offers, pulse width modulation (PWM) and  pulse position modulation (PPM) schemes, in which data are represented by the pulse width and the pulse position, respectively,  have been proposed. The data rate in PWM can be increased by combining it with the discrete multitone technique (DMT) \cite{ntogari2011combining}, while the data rate can be increased in PPM by using overlapping PPM (OPPM) \cite{bai2010joint}, multipulse PPM (MPPM) \cite{sugiyama1989mppm}, or the overlapping MPPM \cite{ohtsuki1993overlapping}. 

Due to the non-linear VLC channel response, the aforementioned modulation schemes suffer from inter-symbol interference. To combat this impairment, the orthogonal frequency division multiplexing (OFDM) scheme, widely used in RF systems, should be modified to be compatible with the IM/DD technique. Because the light signal is a real non-negative signal, the complex bipolar signals generated by OFDM must be represented by real positive signals in VLC. The solution can be implemented by relaxing the Hermitian symmetry constraint and convert the bipolar signal to a unipolar signal.     
Two types of optical-OFDMs are widely used as  VLC modulation schemes: a DC-biased optical OFDM (DCO-OFDM) and an asymmetrically-clipped optical OFDM (ACO-OFDM).  In DCO-OFDM \cite{elgala2007ofdm, afgani2006visible},  a positive direct current is added to make sure that the  signal is non-negative, and all the subcarriers are modulated to maximize the spectral efficiency. On the other hand, in ACO-OFDM, only odd subcarriers are used to modulate the data \cite{armstrong2006power}, resulting in a symmetric time domain signal. 

\subsection{Objectives and Constraints in VLC Networks}

In this section, we present the established objectives for the design or optimization of the VLC networks and discuss the associated constraints that must be achieved. Certainly, some of the unique characteristics of VLC technology  have generated new challenges, different from those in RF networks. As a result, the techniques used in traditional RF networks cannot directly be applied to VLC networks. 

\subsubsection{System Capacity or Sum Rate}

Several issues (that do not exist in the RF systems) must be considered, when deriving the VLC channel capacity. These are: 1) dimming requirements, 2)  peak optical intensity constraint, 3) illumination requirements and the LED dynamic range, 4) and necessity for  the input signal to be non-negative and real-valued. In addition, the channel gain for
VLC is  modeled almost as the Lambertian model \cite{komine2004fundamental}, in which the channel gain for VLC is time-invariant but affected by geometrical parameters such as the locations of the transmitter and receiver. Because of the differences between  RF and VLC systems, the capacity-achieving input distribution does not have to be  Gaussian \cite{lapidoth2009capacity}.  This means that the commonly derived Shannon  channel capacity formula used for RF systems cannot be applied to VLC ones. Consequently, many researchers have been investigating the VLC channel capacity under these constraints.  Several papers have focused on the optical intensity channel capacity, where only two constraints are imposed: the non-negative real-valued intensity signals, and the average light intensity for eye safety standard \cite{lapidoth2009capacity, farid2009capacity, chaaban2016free}. However, for the VLC channel capacity, the illumination requirements and the LED dynamic range constraints must be considered when deriving the channel capacity.  Ahn and Kwon \cite{ ahn2012capacity} proposed a numerical approach to determine the channel capacity for inverse source coding in VLC, without providing a closed-form expression for the VLC channel capacity, whereas Wang \emph{et al.} derived closed-form expressions for the upper and lower bounds of the dimmable VLC channel capacity \cite{wang2013tight}. The lower bound was expressed as follows: 
\begin{equation}
\label{LB}
C \geq \frac{1}{2}\log_2\left( 1+\frac{e}{2\pi} ( \frac{\zeta P}{\sigma})^2 \right),
\end{equation}
where $\zeta$ is the dimming target ranging from 0 to 1, $P$ is the nominal optical intensity of LEDs,  $\sigma^2$ is the Gaussian noise variance, and $e$ is the Euler parameter.     The channel gain (losses and opto-electronic transformation factors) is assumed to be equal to 1, in Equation \ref{LB}.
Expression (\ref{LB}) is the common expression used in the literature to estimate the system capacity. The same authors  derived closed-form expressions for the upper and lower bounds of the dimmable VLC channel, when imposing the peak optical intensity constraint of the LED \cite{wang2014capacity}. This constraint resulted in a loss of  channel capacity, and was found negligible when the maximum allowed optical intensity was twice the nominal optical intensity of the LEDs. With a different approximation method used for the intrinsic volumes of the simplex, Jiang \emph{et al.} \cite{jiang2016tight} derived a tighter upper bound, compared to that derived in \cite{wang2013tight}, for the VLC dimmable channel capacity. 
 Chaaban \emph{et al.}  derived the capacity bounds of the IM/DD optical broadcast channel under two constraints, which are the average light intensity and peak power intensity \cite{chaaban2016capacity}. In \cite{xu2017ergodic}, Xy \emph{et al.} derived a lower bound for the ergodic point-to-point channel capacity. Because the VLC channel is time-invariant and only depends  on geometrical parameters,  the authors derived the ergodic capacity over the communication region in the spatial domain, instead of the time domain. In addition, they derived lower bounds for the ergodic capacity in the dynamic systems for which geometrical parameters follow  typical distributions. For multiple-input-single-output (MISO) VLCs with two users and two transmitters, Agarwal and Mohammed \cite{agarwal2017achievable} proposed an achievable rate region of the proposed system VLC channel when the zero-forcing pre-coder was applied. They showed that the largest rate region was achieved when the average power of LED is half the maximum allowed peak power. 

\subsubsection{Throughput} This criterion is different from the system capacity because it calculates the actual transmitted data rate. Determining the bit error rate (BER) and the used coding and modulation schemes is required in finding the actual system throughput. 

The throughput of the user $j$ can be expressed in OFDM systems using the following expression:

\begin{equation}
X_j=\frac{B}{\beta L} \sum_{i=1}^{\beta L -1} \eta_i,
\end{equation}
where $B$ is the modulation bandwidth, $L$ the number of subcarriers, $\eta$  the subcarrier spectrum efficiency  obtained from the modulation scheme, coding scheme, and the received signal-to-noise ratio (SNR) \cite{burchardt2013distributed}, and $\beta$ is a constant that depends on the kind OFDM used (e.g. for DCO-OFDM $\beta=\frac{1}{2}$).  In TDMA systems, the achieved throughput, at the user $j$, is expressed as follows:
\begin{equation}
X_j=\frac{B}{N_{T,j}} \sum_{i=1}^{N_{T,j}} \eta_i,
\end{equation}
 where $N_{T,j}$ is the number of time slots assigned for user $j$, and $\eta_i$ is the spectrum efficiency of the time slot.
\subsubsection{Energy Efficiency}
 VLC networks are more energy-efficient than RF networks because LEDs, used as transmitters, are energy-efficient devices, and the consumed power used for communication is also used for illumination. However, the range of acceptable illumination values is defined by maximum and minimum requirements, meaning that the consumed power can be controlled, within this given range, to maximize the EE. In addition, with the advent of 5G wireless networks, the tremendous number of access points (APs), and the billions of connected devices, the need for designing energy-efficient
systems is becoming even more compelling for seeking to have green communication systems. 
This is confirmed by what is shown in \cite{TWC} that the EE in VLC networks was greatly affected by an increase in the number of active users. 

The EE can be improved by efficient resources optimization, power allocation, energy transfer and harvesting, and hardware solutions \cite{EEsurvey}. 

The common approach to guarantee energy-efficient systems is to optimally allocate the resources to maximize the EE function subject to QoS and maximum transmit power constraints. The EE function can be defined as the system's benefit over the total consumed power. In other words, if the system's benefit is the sum rate, then the EE is
\begin{equation}
EE= \frac{R_T}{P_T},
\end{equation} 
where $R_T$ is the sum rate and $P_T$ is the total consumed power at the transmitters. 

Another way for improving the EE is to formulate the optimization problem as minimizing the total amount of transmitted power, under a given set of QoS constraints. This type of optimization problems is easier to tackle than the problem of maximization of the EE function. This is because the EE function is not concave, in terms of allocating the transmit power. The common approach to tackle the EE maximization problem is to convert the non-convex problem into a sequence of convex optimization problems using the Dinkelbach's method.  Another way to improve the EE in VLC networks is to harvest the energy by converting the received light intensity into a current used for transmissions. This can be implemented by equipping the receivers with  solar panels.

\subsubsection{Fairness}

 Fairness is an important issue in VLC networks for many reasons: 1) the dramatic decrease in the VLC channel value with the distance between the transmitter and receiver makes many users unable to switch from crowded cells to uncrowded ones; 2) the small coverage stimulates designers to fully re-use the frequency in the cells, resulting in  severe interference with the signal received by some users. 
 
 
 Fairness is commonly  measured using Jain's formula \cite{jain1984quantitative} for a single cell or for the whole cellular system. The cell fairness is 
\begin{equation}
F_i=\frac{(\sum_{j=1}^{N_i}R_{j,i})^2}{N_i\sum_{j=1}^{N_i}R_{j,i}^2},
\end{equation} 
and the fairness of the whole cellular system is
\begin{equation}
F_s=\frac{(\sum_{i=1}^{N_{ap}}\sum_{j=1}^{N_i}R_{j,i})^2}{N_{ap}\sum_{i=1}^{N_{ap}}\sum_{j=1}^{N_i}R_{j,i}^2},
\end{equation}
 where $N_i,\ N_{ap}$ are the number of users associated to the cell $i$ and the number of cells, respectively, and $R_{j,i}$ is the $j$th user data rate associated with the cell $i$.

 Fairness can be achieved either by formulating the optimization problem to maximize the utility with a guarantee to achieve a proportional fairness \cite{li2008proportional},  $\alpha$-proportional fairness \cite{uchida2009information}, or by adding the QoS constraints to the formulated optimization problem. The concept of the proportional fairness is to modify the objective function to imply both the system utility (e.g. sum-rate) and the fairness. If we denote $x$ as the utility function, the generalized objective function can be written as follows:
 
 \begin{equation}
 \Gamma(x)= \begin{cases} \log(x), & \alpha=1;\\
 \frac{x^{1-\alpha}}{1-\alpha}, &\alpha\geq0, \alpha \neq 1,
 \end{cases}
 \end{equation} 
where $\alpha$ is a proportion factor. $\alpha=0$ means the utility is only considered and the fairness is ignored;  $\alpha=1$ means that the proportional fairness is achieved, and, if $\alpha \rightarrow \infty $, the max-min fairness is achieved.  

\subsubsection{Required Illumination Constraints}
\label{RIC}

The two functions of LED, illumination and communication, are related to each other and must be studied and optimized jointly. In other words, the illumination requirements should be considered in designing the input current to the transmitter LED.  This requirement implies that different constraints must be considered when optimizing the communication in VLC networks. The constraints are the peak optical power,  dimming requirements, and  flicker reduction.   

For the peak power constraint, we should note that the input signal to the LED contains two components: the alternative signal (that contains the information), and the DC signal used to guarantee non-negative signal. The total energy emitted by the LED determines the transmitted optical power and the subsequent received signal strength, whereas the brightness is determined by the luminous intensity \cite{komine2004fundamental}. We denote $\Phi_{max},\ \Phi_{min}$, and $\Phi_{avg}$, as the predefined minimum illumination, maximum illumination, and the average illumination  over the entire area, respectively.  For the office work, an illuminance between 300 to 2500 lux is required \cite{komine2004fundamental}.

The relation between the radiated optical power at LED and the luminous flux at the point $i$, which is distant from LED by $d_i$ m with incidence and radiance angles $\theta$ and $\psi$, respectively, can be given by \cite{hanzo_haas, grubor2008broadband}   
\begin{equation}
 h_i P_{opt} = \delta \Phi_i,
\end{equation}
where $\delta$ is the optical to luminous flux conversion factor \cite{grubor2008broadband} which its value depends on the LED type;  $P_{opt}$ is the optical power, $\Phi_i$ is the luminous flux at point $i$, and $h_i$ is given by:

\begin{equation}
h_i=\frac{m+1}{2 \pi d_i^2} cos^m(\theta)cos(\psi),
\end{equation} 
where $m$ is the Lambertian index given in Section \ref{CM}.


One additional constraint for communication is that the input DC-biased current (DC and AC currents) to the LED must be within the dynamic range of the LED to have the radiated optical power linearly proportional to the input current \cite{dimitrov2012clipping}. For instance, the practical dynamic range of the LED Vishy TSHG8200 is within [5 mW, 50 mW]. 



To meet the illumination requirements at all points in the floor area, the upper and lower bounds of the optical power should be set accordingly. Considering both the bounds of the LED dynamic range and the illumination limits, the optical power at the transmitting LED must be confined by 

\begin{equation}
\max(P_{min,ill}, P_{min,D}) \leq P_{opt} \leq \min(P_{max,ill}, P_{max,D}),
\end{equation} 
where $P_{min,ill}$ and  $P_{max,ill}$ are the minimum and maximum optical power required for achieving the corresponding illumination requirements, respectively; $P_{min,D}$ and $P_{max,D}$ are the maximum and minimum power limits for the LED dynamic range, respectively.


The dimming control is a desirable process for the illumination purpose \cite{gancarz2013impact}.  For power saving, LEDs can be dimmed to desired levels, using  appropriate modulation schemes \cite{rajagopal2012ieee}, such as multi-pulse position modulation (M-PPM) \cite{lou2017joint}; or variable  OOK \cite{lee2011modulations}.  

Another purpose for the used modulation scheme is to mitigate the light intensity fluctuation to be unnoticeable by the human eyes. To guarantee the flickering is above the human eyes' fusion frequency, flickering frequency must be at least greater than 200 Hz \cite{ieee_std};  this can be avoided by using the Run Length Limited (RLL) codes that are used to reduce the long runs of 0s and 1s.

\subsubsection{Coverage Probability}

\begin{figure*} 
    \centering
  \subfloat[FoV = 30$^o$]{
       \includegraphics[width=0.49\linewidth]{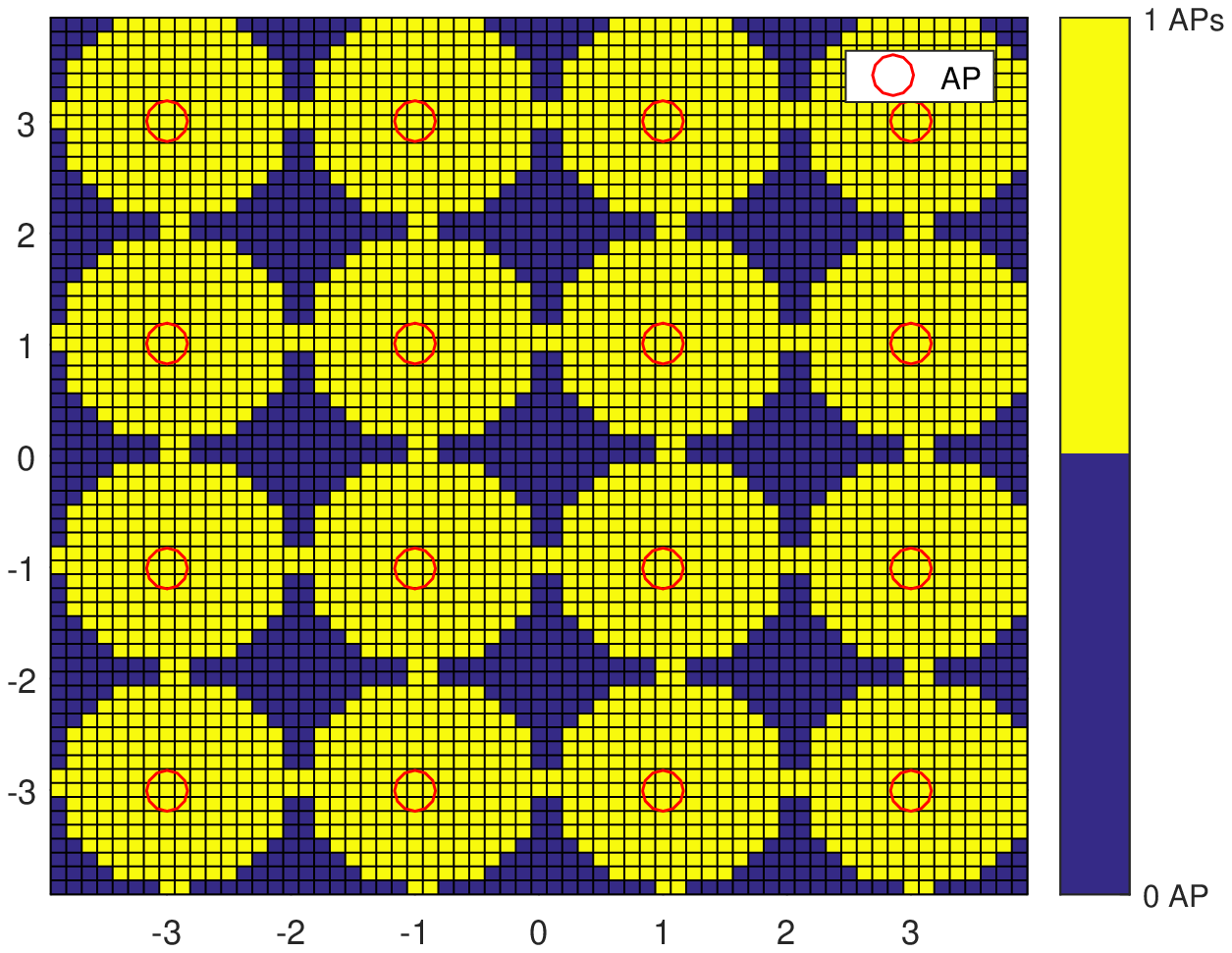}}
    \label{1a}\hfill
  \subfloat[FoV = 40$^o$]{
        \includegraphics[width=0.49\linewidth]{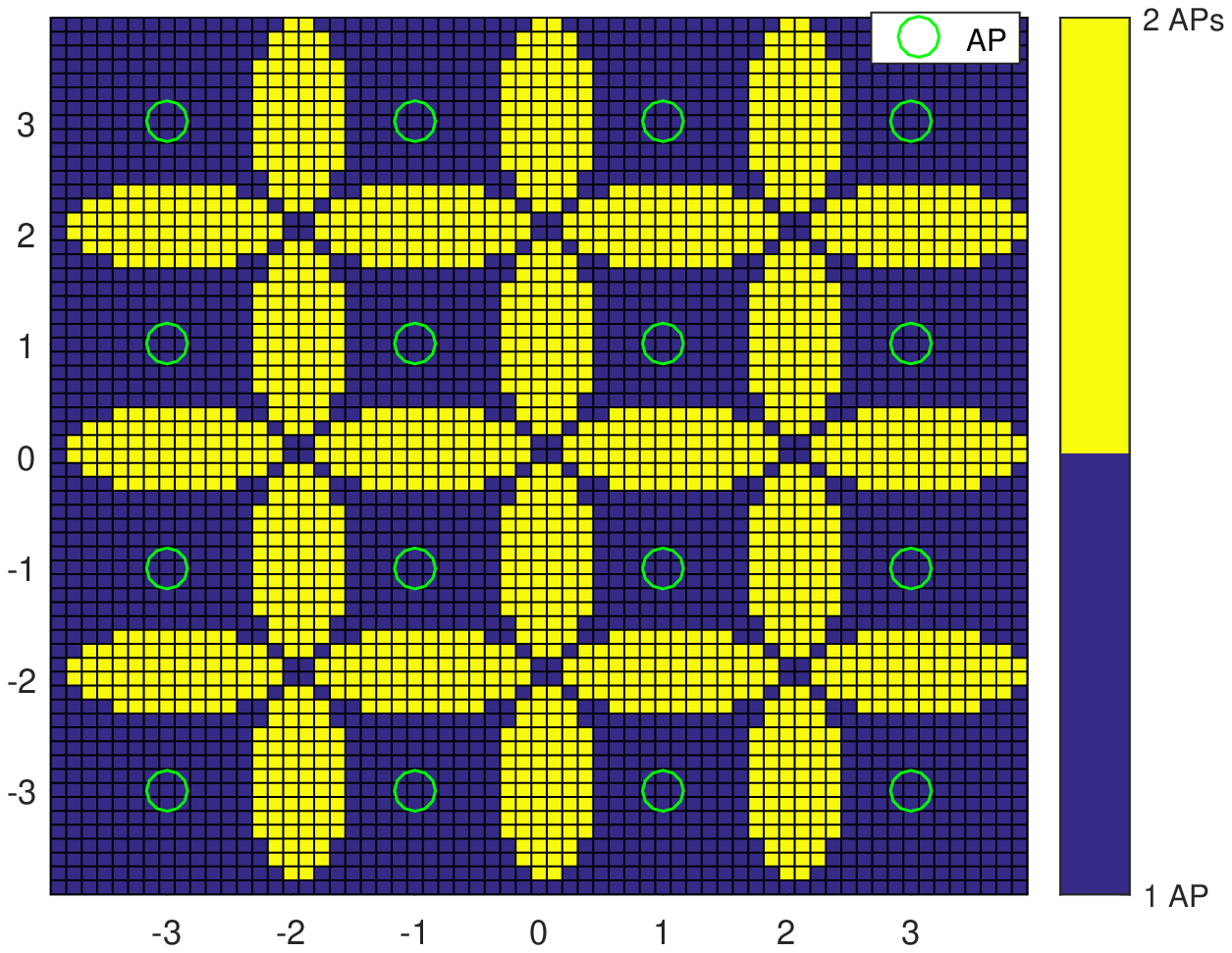}}
    \label{1b}\\
  \subfloat[FoV = 50$^o$]{
        \includegraphics[width=0.49\linewidth]{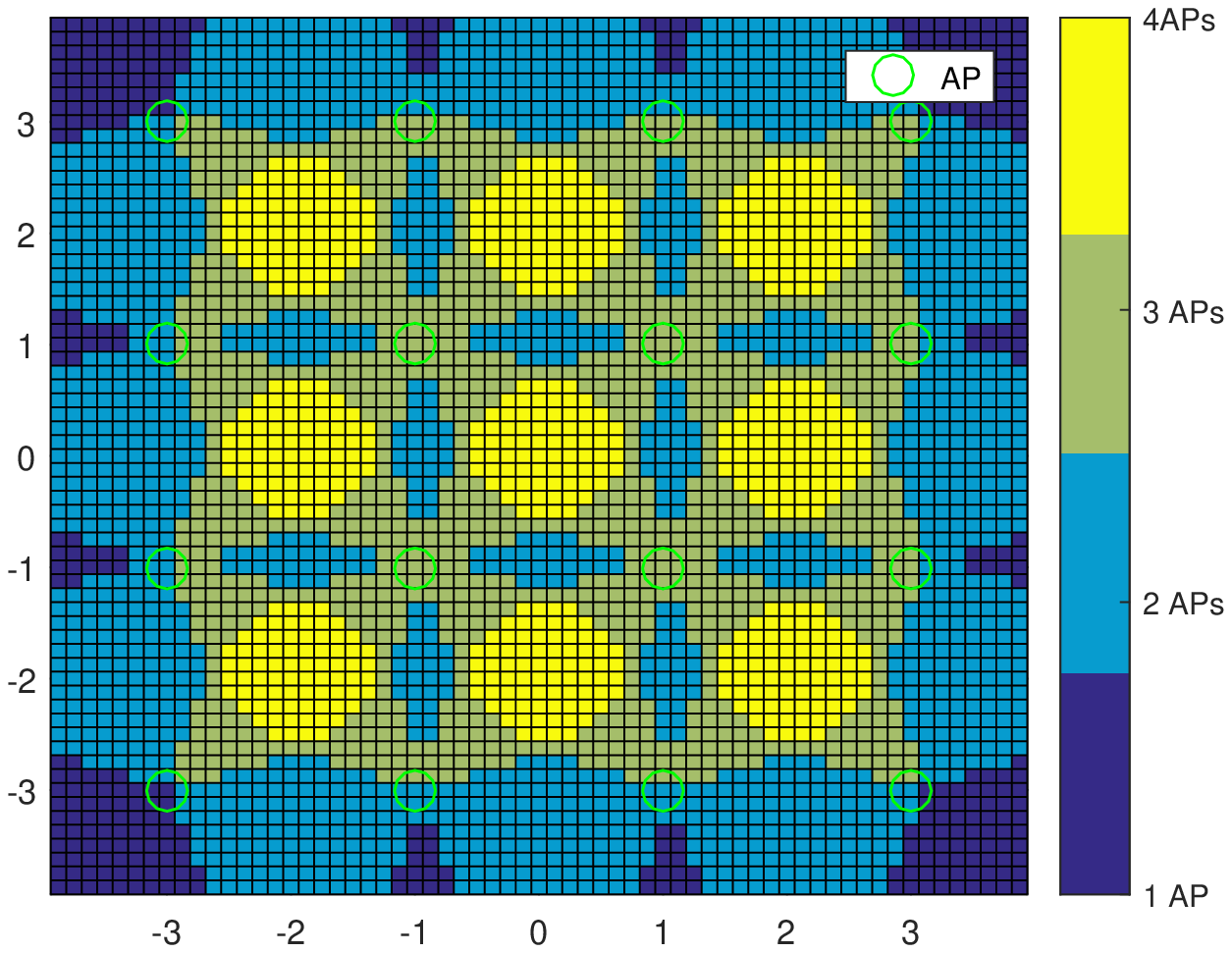}}
    \label{1c}\hfill
  \subfloat[FoV = 60$^o$]{
        \includegraphics[width=0.49\linewidth]{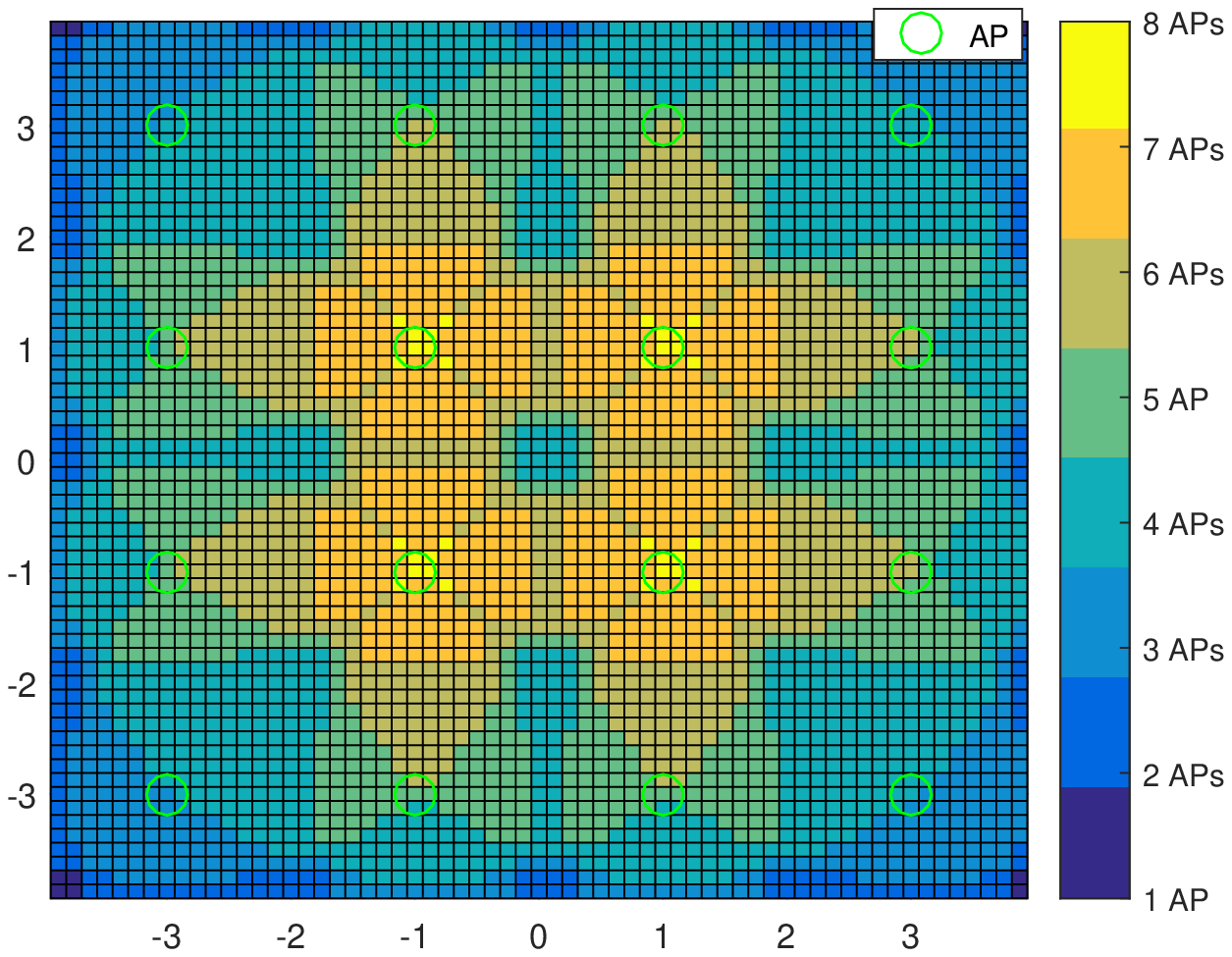}}
     \label{1d} 
  \caption{Number of APs that can cover the area based on the user's FoV.}
  \label{coverage} 
\end{figure*}

\begin{table}[!htbp]
\centering
\caption{Simulation Parameters}
\label{table}  
\begin{tabular}{|p{.28\textwidth} | p{.11\textwidth} |}
\hline
  Name of the Parameter & Value of the Parameter\\
 \hline 
 
  Maximum bandwidth of VLC AP, $B$ & $20$ MHz  \\
  
  The physical area of a PD for IUs, $A_{p}$ & $0.1$\ cm$^2$ \\
  The physical area of a PD for EH users, $A_{p}$ & $0.04$\ m$^2$ \\
   Half-intensity radiation angle, $\theta_{1/2}$ & $60^o$\  \\
   FoV semi-angle of PD, $\Theta$  & $30^o-60^o$\\
  Gain of optical filter, $g_{of}$ & $1$  \\
  Refractive index, $n$ & 1.5 \\
   Efficiency of converting optical to electric, $\rho$& $0.53$ [A/W]\\
  Maximum input bias current, $I_H$ & $12$ mA  \\
  
  Minimum input bias current, $I_L$ & $0$ A  \\
  Fill factor, $f$ &0.75\\
  LEDs' power, $P_{opt}$ & 10 W/A\\
  
  Thermal voltage, $V_t$ & 25 mV \\
  
  Dark saturation current of the PD, $I_0$ & $10^{-10}$ A\\
  
  Noise power spectral density of LiFi, $N_0$ & $10^{-21}$\ A$^2$/Hz  \\
  Room size, & $8\times8$\\
  Room height, &$2.5$ m\\
  User height & $0.85$\\
  Number of VLC APs, &$ 4\times4$\\
  Number of users, & 5-35\\ 
  Monte-Carlo for user distribution,  & 100 different user distributions\\

  \hline 
  
  RF    \\
  \hline
  Number of RF APs & 1\\
  Location of RF AP & (0,0) in the ceiling\\
   Transmit power & 10 Watt\\
  The distance of breakpoint & 5 m\\
  Central carrier frequency  & 2.4 GHz\\
  Bandwidth & 20 MHz\\
  Angle of arrival/departure of LoS & 45$^o$\\
  Standard deviation of shadow fading (before the breakpoint) & 3 dB  \\
  Standard deviation of shadow fading (after the breakpoint) & 5 dB  \\
  Noise power spectral density & -174 dBm/Hz\\
  \hline  
\end{tabular} 

 \end{table}

Since the LEDs in VLC can cover only a small area, and the coverage probability decreases dramatically as the distance increases, the coverage is an important issue in VLC networks and should be considered when designing the networks' parameters. The coverage probability can be defined as the probability that the received data rate for typical user is greater than or equal to a certain data rate threshold. All the geometrical parameters of the VLC channel affect the coverage probability, but we focus our discussion on those having major impacts such as the distance, optical power intensity, and the user's field-of-view (FoV). If we consider a system model consisting of multiple VLC APs and the considered user $j$ is served only by one AP $i$, increasing the optical power would surely enhance the channel link from the AP $i$ to the user $j$, but would increase the interference from all other APs significantly.   The user's FoV plays a significant role in affecting the coverage probability, since decreasing the user's FoV leads to enhancing the VLC channel and decreasing the number of interfering APs, but we should also note that an extensive decrease in the user's FoV leads to decrease of the coverage probability. On the other hand, for a given  FoV, increasing the height of the APs leads to an increase in the number of APs  in the user's field of view, meaning that the number of  interfering APs would increase, and the path loss from the AP $i$ to the user would also increase. 

\begin{figure}[t]
\centering
\includegraphics[width=3.5in]{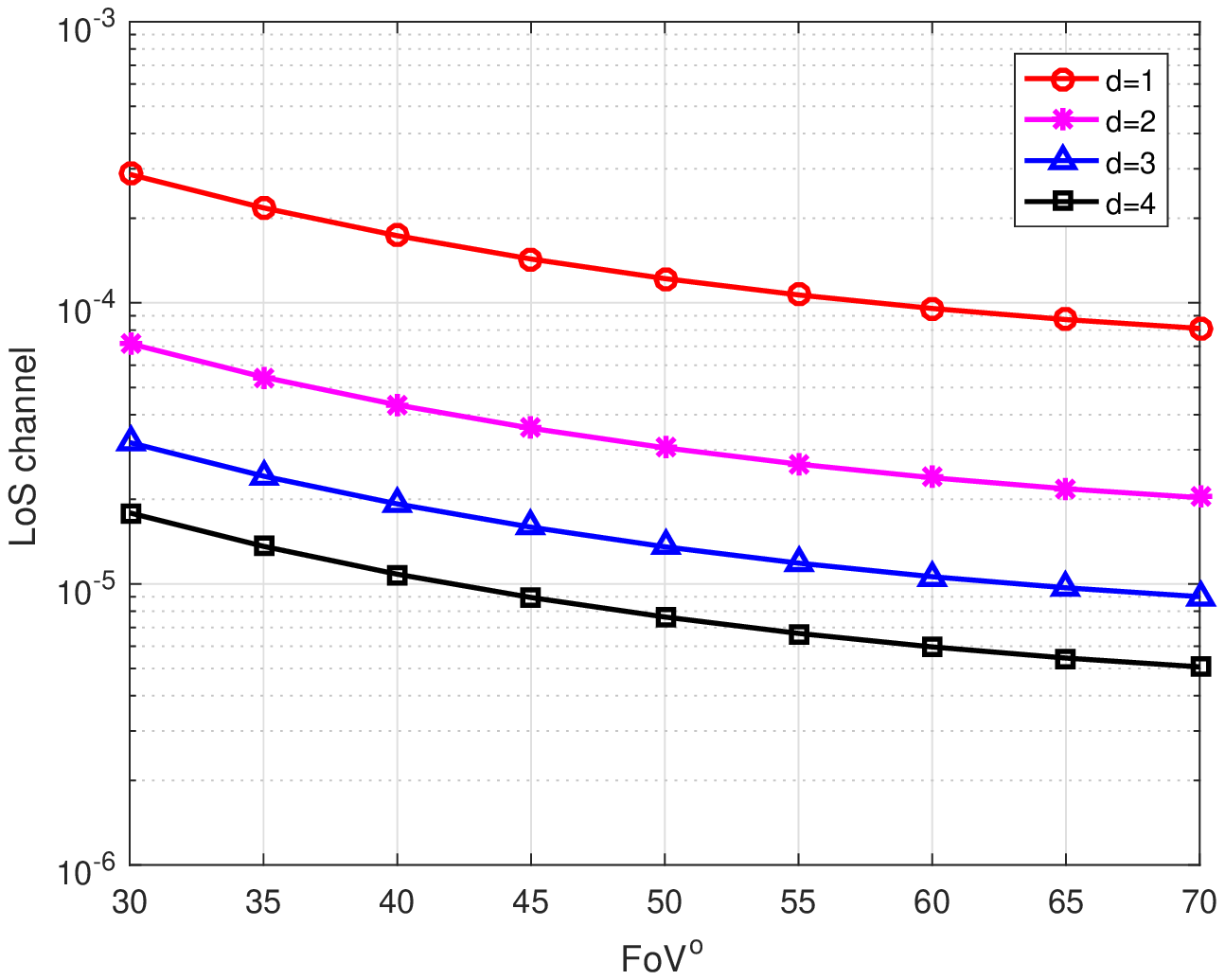}
\caption{The effect of user's FoV on the channel quality with different transmitter-receiver distance, when the angels of radiance and incidence are zero.}
\label{LoSvFoV}
\end{figure}

Fig. \ref{coverage} represents  the effect of a user's FoV on the coverage probability by showing the number of APs that can cover the area with different user's FoV. Fig. \ref{coverage} shows that the coverage probability increases as the user's FoV increases. On the other hand, the channel quality decreases as the user's FoV increases (Fig. \ref{LoSvFoV}). Both figures show that the user's FoV has a great impact on the channel quality and the coverage probability, meaning that optimizing the  FoV would  have a significant impact on the VLC systems.  Table \ref{table} contains the simulation parameters considered in our study.

\subsubsection{The Harvested Energy}

An additional function to LEDs, besides the illumination and communication, is the transfer of power, using the light intensity. When the VLC network consists of users that need to harvest the energy, the parameters should be designed to find a compromise between the three functions. The receiver can harvest the energy by equipping it with a solar panel that can convert the received modulated light signal into an electrical signal without an external power supply. Because the received current signal at the receiver contains both DC and AC currents, the DC current can be blocked and forwarded to the energy harvesting circuit.   Li \emph{et al.}, in \cite{solar}, derived the energy that can be harvested by a user from one LED as: 

\begin{equation}
E=fI_{DC}V_{oc},
\end{equation}
where $f$ is a fill factor of approximately 0.75, $I_{DC}$ the received DC current, and
\begin{equation}
V_{oc}= V_t \ln(1+\frac{I_{DC}}{I_0}),
\end{equation}
where $V_t$ is the thermal voltage, and $I_0$  the dark saturation current of the PD. If we denote the transmitted DC current by $b$, the received DC current can be expressed as by $I_{DC}=\rho h P_{opt}$. Hence, if we have multiple LEDs, the harvested energy at the user $j$ is given by:
\begin{equation}
\label{EH}
E_j=f \rho P_{opt} V_t  \mathbf{h}_{j}^T\mathbf{b}\ln(1+\frac{\rho \mathbf{h}_{j}^TP_{opt}\mathbf{b}}{I_0}),
\end{equation}
where $\mathbf{h}_j$ is the channel vector between the LEDs and the user $j$, and $\mathbf{b}$ is the DC Bias current vector at LEDs.  
\subsubsection{Secrecy Capacity}

When VLC networks contain two types of users, the authorized users that have the authority to obtain and decode the data, and the eavesdroppers trying to obtain confidential messages without permission, the performance metric is changed to be the secrecy capacity. The secrecy capacity is defined as the maximum information rate that can be attained by the legitimate receiver minus the maximum eavesdropper's information rate\cite{liang2009information}. If  the average power constraint is only considered in the Gaussian wiretap channel, the optimal input distribution is Gaussian; however, if the amplitude power constraint is considered, it is difficult to find the optimal input distribution for capacity-achieving \cite{mostafa2016optimal}, but the lower and upper bounds can be found.  However, for the uniform input distribution and with considering amplitude constraint $\vert x(t) \vert \leq A \ \forall t$, the secrecy capacity of the single-input-single-output (SISO)-VLC system is lower and upper bounded, respectively, by \cite {mostafa2015physical2}

\begin{equation}
C\geq \frac{1}{2} \log \frac{6h^2_D A^2+3 \pi e \sigma^2}{\pi e h_E^2 A^2+3 \pi e \sigma^2},
\end{equation} 
and 
\begin{equation}
C\leq \frac{1}{2} \log \frac{h^2_D A^2+\sigma^2}{ h_E^2 A^2+ \sigma^2},
\end{equation} 
where $h_D$ is the transmitter-legitimate receiver channel, $h_E$ is the transmitter-eavesdropper channel, $\sigma$ is the noise variance, $e$ is the Euler parameter, and $A$ is the peak power constraint. 


\section{Resource and Power Control with AP Assignment}

In this section, we review the optimization techniques previously reported in the literature  to improve the VLC network performance when the system consists of multi-users. Four main issues are considered in this type of networks, for maximizing the various objectives and achieving the various constraints discussed in Section II. These are: the user-to-network association (called 'access point assignment' (APS)), resource management, power allocation, and APs coordination. The joint of APA and resource allocation was identified by load balancing (LB). LB has been extensively investigated in RF networks \cite{andrews2014overview, aryafar2013rat}. However, the unique properties of VLC technology make the problem different, and the techniques used in RF networks cannot be  directly  applied to VLC networks.  

\begin{figure}[t]
\centering
\includegraphics[width=3.3in]{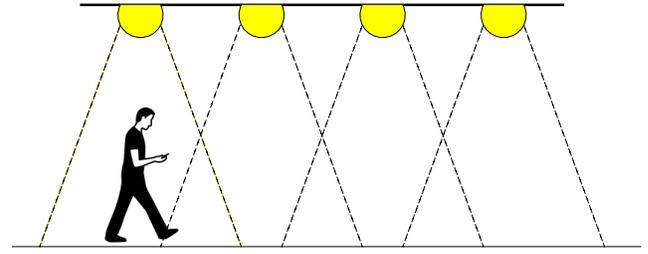}
\caption{Handover in VLC network.}
\label{handover}
\end{figure}

Despite all the advantages of VLC systems mentioned in the Introduction, they suffer from several limitations that contribute to the degradation of the system's performance such as  a small coverage area, non-LoS failure transmission, frequent handover, and inter-cell interference. This leads to unbalanced systems,  with some users receiving a poor service, while others may receive a high QoS. For instance, the opaque placed in the indoor environment might block the LoS light that carries data for some intended users, leading to a degradation of the channel by up to 90 percent of the LoS channel \cite{chen2015fractional}, and, as a consequence, a significant deterioration of the data rates for the intended users. However, these opaque objects can block the inter-cell interference coming from the adjacent VLC APs for other users. This means that the fluctuation of the received QoS at users is high and that the blockages significantly affect the system fairness and the balance of the systems.  Another cause for unbalanced VLC systems is the handover. For the reason that the coverage area of LEDs is small, the mobile users would suffer from wasting resources by sending and transmitting the overhead of the required handover.   Fig. \ref{handover} shows how the handover is an important issue in VLC systems. As observed in the figure, the small coverage area of the LEDs in  VLC networks leads to a decrease in the throughput of  both the system and the mobile users due to the overhead generated by such handovers \cite{hand, mobility, dyn}. However, by dividing the time into  sufficiently short periods, we can have  quasi-static periods known as  'states'.  The handover consumes time, on average from 30 ms to 300 ms \cite{fast2008}. Another issue due to the small coverage area is the fact that the crowded static users cannot be distributed to the deployed cells, resulting all or most of them will be connected to one cell. This causes some APs to be overloaded, and consequently leads to a poor service for the connected users, while the other APs are unloaded or have a lower number of users.  The bright side of the VLC's small coverage area is the fact that the whole bandwidth can be fully re-used in all cells, which improves the spectral efficiency of the overall system \cite{haas2016lifi}. However, re-using the full frequency in cells generates inter-cell interference, to some extent. Inter-cell interference can be accepted for the sake of improving the system's spectral efficiency. On the other hand, the services received by the users located at the edges of the cells would be affected by this inter-cell interference. To summarize, because of these issues, the users located at the edges of cells, blocked by objects, in motion, or connected to overloaded APs can not receive a good QoS like the other users. This significantly deteriorates both the performance and fairness of the VLC systems.

\subsection{Optimizing Hybrid VLC/RF Networks}
\label{hybrid}

\begin{figure}[t]
\centering
\includegraphics[width=3.5in]{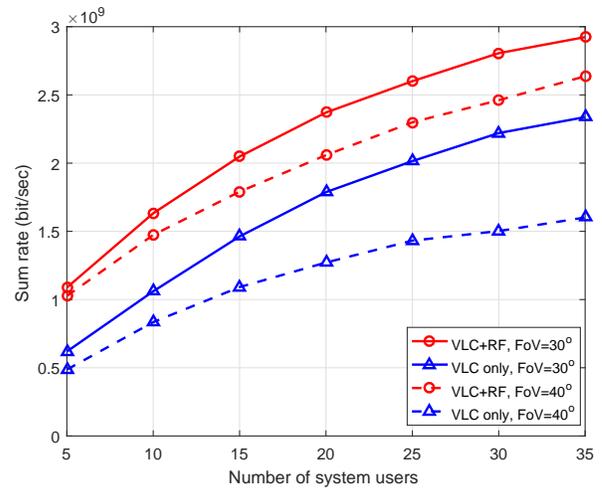}
\caption{Comparison of VLC/RF system and VLC alone by plotting the sum rate versus the number of system users with different users' FoV.}
\label{SRvNu}
\end{figure}

\begin{figure}[t]
\centering
\includegraphics[width=3.5in]{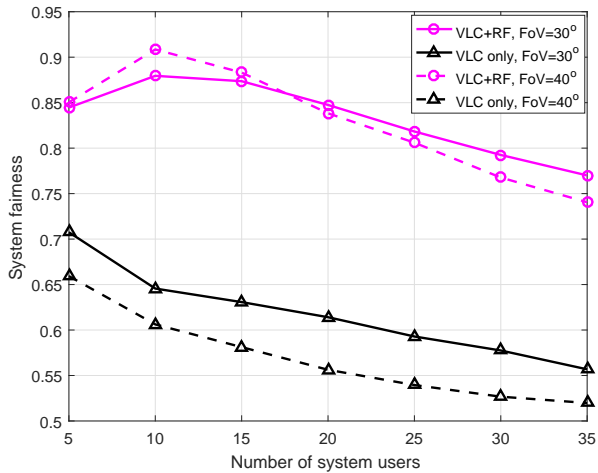}
\caption{Comparison of VLC/RF system and VLC standalone by plotting the system fairness versus the number of system users with different users' FoV.}
\label{SFvNu}
\end{figure}

 One of the most common solutions to the aforementioned VLC issues is to supplement the standalone VLC networks with RF networks. Compared to VLC networks,  RF networks are known for their ubiquitous presence (high coverage area) and proper operation in  non-LoS environments. In addition, the  devices connected to RF networks do not suffer from VLC interference and vice-versa \cite{ayyash2016coexistence}.  Therefore, adding one or more RF APs to VLC networks mitigates the LoS blockages, handover overhead, and inter-cell interference. 
  However, a problem remains: finding  a compromise between the high coverage area RF networks and the high capacity VLC networks. In other words, how to distribute the users among the APs (either RF or VLC) to improve the overall system's performance with an acceptable fairness of the system. The main idea is to associate the users who suffer from interference, handover overhead, and blockages to the RF AP(s) and keep the other users connected to the VLC networks. As shown in Fig. \ref{coverage}, when the users' FoV is 30$^o$, the problem in VLC networks is the coverage, whereas, if the users' FoV is greater than, or equal to 40$^o$, the problem is the interference. In Figures \ref{SRvNu} and \ref{SFvNu}, we show how adding one RF AP to the VLC network can enhance the sum rate and system's fairness, respectively. In these figures, we associate the uncovered users (when FoV = 30$^o$) and the interfered users (when FoV = 40$^o$) to the RF AP, while keeping the other users connected to the VLC network.  The simulation parameters are shown in Table \ref{table}, and the RF channel is modeled as in \cite{access2017}.

  Several techniques have been proposed  to balance the load and tackle these issues by an efficient user distribution among VLC/RF APs  \cite{hybrid2014, improve,dyn,coop,hand,wang2017optimization, wu2017joint,wu2017access, stage, access2017, wang2016fuzzy,shadow,comparison2017,learning2017,mobility,globecom,JOCN}.  LB consists of two missions: the APs' assignment (APA) and allocating the resources, whether this resource is a time slot in TDMA schemes or a sub-carrier in OFDMA schemes. Specifically,  Stefan and Haas \cite{hybrid2014} started to study the APA by distributing the users between one RF AP and one VLC AP. Some of the users were associated to the VLC AP to alleviate the load of the RF AP, and the infeasible VLC connections were transferred to the RF AP.  In \cite{improve}, by having multiple VLC  and RF APs, the advantages of combining 
  RF and VLC networks were investigated, and it was proposing that users can be distributed dynamically, on both the VLC and RF networks, based on the users' channel condition. Users can then migrate to the AP offering higher data rates. The APA was implemented in \cite{improve} under the assumption that the resources are allocated fairly among users. It was concluded that the hybrid VLC/RF networks improved the performance significantly, compared to the VLC or RF standalone networks. Authors of \cite{dyn} proposed to first associate the users to the VLC network, and then, to re-allocate the users receiving a lower data rate than a predefined threshold to RF APs. In \cite{coop}, authors formulated a centralized and distributed optimization problem for user association to the APs (whether this AP is VLC or RF AP) with allocating the resources jointly among users. The centralized optimization problem, with considering the proportional fairness \cite{kelly1997charging}, was formulated as a mixed-
integer non-linear programming (MINLP), which is highly complex. Hence, a distributed algorithm was also proposed with lower complexity compared to the centralized algorithm. 

To decrease the number of handovers,  Wang and Haas \cite{hand} proposed a dynamic LB scheme in which the quasi-static users are
connected to   VLC APs, and the moving users are connected to the RF
AP. In \cite{ wang2017optimization, wu2017joint}, authors upgraded the formulated optimization problem in \cite{coop} to consider the handover in the dynamic systems. 
With considering the handover overhead and $\alpha$-proportional fairness, the authors of \cite{wang2017optimization} formulated and proposed two solutions for two optimization problems, i.e. the joint APA and resource allocation problem (JOA), and the separate APA and resource allocation (SOA).  They compared  the two approaches in terms of performance and complexity. The former approach was found to achieve a better QoS for the users, but with a significant higher complexity, up to 1000 times greater, than the later. In a separate study \cite{ wu2017joint},  instead of assigning the users to a specific AP, Wu \emph{et al.} formulated the problem by considering the handover as a hierarchal assignment to first assign the network (either RF or VLC) to each user, and then select the appropriate AP, in the assigned network, for each user. Because the problem formulated in \cite{coop} is for static systems, those presented in \cite{ wang2017optimization, wu2017joint} provide a significant improvement in the system performance for dynamic systems. 

Instead of  considering the handover with LB, Wu and Haas \cite{wu2017access} considered the LoS VLC channel blockages in the formulated optimization problem. They  modified the formulated optimization problem to accommodate the LoS VLC channel blockages. The main idea is that, the users that suffer from a high occurrence rate of channel blockages should travel to the RF networks, whereas the users that do not suffer from  blockages, or the ones that suffer from a low rate of blockages (to avoid the effect of handover overhead), should stay in the LiFi networks.

 To avoid the complexity of solving these optimization problems, fuzzy logic-based approaches were proposed for balancing the load in VLC networks \cite{stage, access2017}, and \cite{wang2016fuzzy}. Authors of \cite{stage} and \cite{access2017} proposed two-stage assignment process for the users in one RF AP and multiple VLC APs. They first decided which users should be connected to the RF AP, then they distributed the remaining users to the VLC APs, regardless of the presence of the RF AP and its connected users. In the fuzzy logic approach, the user $j$ scores the APs, based on its offered throughput,  SNR, inter-cell interference from the adjacent  APs,  and  activity of the adjacent VLC APs, then decides whether to connect to the RF AP or to the VLC network, based on the resulting score.  Similarly, authors of \cite{ wang2016fuzzy} used this approach to handle the handover in a dynamic hybrid VLC/RF system model. In their scheme, they considered several parameters as an input to the fuzzy logic approach: the instantaneous and average CSI, user speed, and the minimum required data rate at users.
 
 In \cite{shadow}, authors used another approach called  the 'evolutionary game theory' (EGT), to solve the joint LB and resource allocation problem. Some  practical issues were considered in their study, including the receiver's orientation angle, LoS blockage in RF and VLC APs, and the diversity in the users' data requirements. In addition, the channel of LiFi was characterized with considering these practical factors.    Authors in \cite{comparison2017} studied and compared the common approaches used for balancing the load in the hybrid VLC/RF networks which are: 1) optimization based algorithms, 2) evolutional game theory, 3) fuzzy logic based algorithms. They showed that, for the dynamic systems when the handover is considered besides the AP assignment and the resource allocation, the fuzzy-logic-based algorithms outperformed the other approaches, whereas for the static systems, the optimization-based algorithms are the best, with a slight improvement over the simpler EGT approach. 
 
 Authors of \cite{learning2017} used a different approach for assigning the APs in the dynamic systems, using bandit theory with considering the accumulated reward gap function as a performance metric. Their idea was to consider the learning aided AP assignment that enables the system to adjust the AP selection probability depending on the learning historical reward information and the environmental information. In \cite{mobility}, authors distributed  users to the APs by applying the matching theory, as the users were mapped to be students and the APs were mapped to be collages. Then, taking into account the preferences (i.e. system throughput, users' moving directions, and fairness index),  students (users) would decide which  collage (AP) is the best for them to maximize their preferences, in return,  collages accept the maximum number of applicants to maximize their preferences, while rejecting the others. The rejected students  would go to their second preferable collage (AP), and so on. 
 
 In a different way, we proposed new algorithms for  joint APA and power allocation aiming to improve both the system capacity and fairness \cite{globecom, JOCN}. Because the assignment of APs, power allocation, and determination of the exact interference information are interlinked problems,  iterative algorithms were proposed to efficiently jointly distribute the users to APs, and to distribute the powers of the APs to the users.  

Some studies focused on allocating the resources, rather than APA \cite{jin2015resource, kafafy2018novel,Energy, kafafy2017power,kashef2017transmit}. These methods are appropriate for the quasi-static systems and when the LoS blockages are not present. In \cite{jin2015resource}, authors considered both the multi-homing and multi-mode mechanisms and they formulated for each mechanism an optimization problem to allocate the resources for maximizing the effective capacity by satisfying the statistical delay target. In the multi-homing mechanism, users can gather the information from the VLC and RF APs at the same time, whereas in a multi-mode mechanism,  users can be connected to one type of networks only. Unlike the multi-homing mechanism, the centralized formulated optimization problem for multi-mode mechanism needs to select the AP for each user; therefore,  a computationally intractable approach was considered in \cite{jin2015resource}, and a distributed suboptimal method was proposed. They showed that, by tightening delay requirements, the multi-homing mechanism provides a much better performance. In \cite{kafafy2018novel}, for multi-users with multiple VLC APs and one RF AP, authors studied the problem of maximizing the EE under maximum power constraints on both RF and VLC APs, and under QoS constraints, when the multi-homing mechanism was applied. 
In \cite{Energy}, authors expanded on the work presented in \cite{kafafy2018novel} and \cite{kafafy2017power}, and jointly allocated the power and  bandwidth to the users, but in only one VLC and one RF AP.
Both \cite{kafafy2018novel} and  \cite{Energy} used Dinkelbach method to convert the nonconvex problem to a sequence of convex problems, then used the sub-gradient method to solve those convex problems. By assuming that a multi-homing mechanism is available to users, there is no need to balance the load by efficiently distributing the users between the RF and  VLC APs. In \cite{ kashef2017transmit}, a system consisting of a cascaded power-line-communication (PLC)/VLC link, along with a RF link was optimized, meaning that the total transmitted power under QoS constraints was minimized. The formulated optimization problem was shown to be a convex problem that could be solved efficiently. In \cite{Zhang2018EE}, authors  formulated a power and sub-channel allocation optimization problem for energy-efficient software-defined  VLC/RF  network, when the users have the multi-homing capability. The optimization problem considered the backhaul constraints, QoS requirements, and the inter-cell interference constraints. With the help of the software-defined controller, the resource allocation strategy can be requested as an application from the application layer, then through the software-defined controller, the requested strategy can be implemented in  the APs in the physical layer.  Because the objective function is the nonconvex EE function, the Dinckelbach approach was also used to convert the problem into a serial of convex optimization problems. 

In \cite{ design2017}, a comparison between the performance of the standalone VLC networks with that obtained from augmenting RF APs to the VLC network (in terms of outage probability) was provided. Specifically, authors quantified the minimum required RF resources (bandwidth and power) for the VLC networks to achieve a predefined (per user) rate outage performance. In \cite{coverage2018}, Tabassum and Hossain used the stochastic geometry to analyze the coverage and the rate of a typical user and  compared the results in four types of networks: RF-only, VLC-only, opportunistic RF/VLC (either the user connected to RF or VLC), and the hybrid RF/VLC (the user can gain the resources from both the RF and VLC APs) networks.  Based on several parameters including the FoV receiver, number of interfering LEDs,  distribution of the interference, association and coverage probability, and the average rate of the typical users, they found closed-form solutions to distribute the users among the VLC and RF APs. By imposing the QoS constraints based on the data link metrics, i.e. the limits on the buffer overflow and buffering delay probabilities, Hammouda \emph{et al.} \cite{ hammouda2018link} showed that the VLC links offered  queuing delays lower than RF links when the data arrival rates at the transmitter buffer were low; however, the RF links supported the higher data arrival rates. 

\subsection{Optimizing the Standalone VLC Networks}

As previously shown in \ref{hybrid}, the most common solution for handover, LoS blockages, coverage, and the inter-cell interference is to support the VLC network by a RF network. However, some studies reported in the literature focused on the LB in standalone VLC networks. 

In \cite{ soltani2016access}, with the help of a central controller, and by considering the arbitrary receiver orientation, Soltani \emph{et al.} proposed an approach for APA to users, based on the strength of the received signal and the traffic of the APs, aimed at maximizing the system's throughput. Briefly, when a coming user wants to join  an established network, the central controller calculates all the offered data rates from all APs and enables the user to select the best AP for him. In \cite{BAG}, authors jointly allocated time resources to the users and assigned APs to the users. They conceived the problem as a bidirectional allocation game, since the aim of APs is to select the only users that maximize the system throughput, and the users want to select  APs providing better QoS. By  considering mobile users in standalone VLC networks, Zhang \emph{et al.} \cite{zhang2018anti} proposed a novel user-to-AP assignment based on anticipating the future users’ locations and their traffic dynamics, and find a trade-off between the delay and the throughput in the dynamic VLC systems. In \cite{ jiang2017joint}, authors studied and formulated the joint power allocation and LB problems. By considering a proportional fairness \cite{ kelly1997charging},  the formulated optimization problem was found an intractable nonconvex. Thus,  a suboptimal solution was proposed to optimize both the power and the time fraction in an alternating fashion.

Another factor that can be used to enhance the performance of VLC networks is the arrangement of APs, in which  APs are placed and selected in the most appropriate way to improve both the illumination and communication. In \cite{ chen2016downlink}, authors investigated the effects of the cell size and network deployment on the performance of VLC systems by measuring the signal-to-noise and interference ratio (SINR) distributions, outage probabilities, and data rates. They concluded that the hexagonal cell deployment achieved the best performance, whereas the random cell deployment exhibited the worst performance. In addition, they demonstrated that the multipath effect was much less prominent than the effect of the co-channel interference because of the PD's size compared to the light wavelength. They also compared the performance of the VLC with the RF and mmWave indoor networks and showed the superiority, in general, of the VLC systems. 
 
 The aforementioned papers optimized the VLC networks based on a TDMA scheme. In \cite{wang2017ofdma} and \cite{ling2018efficient},  the resources in an OFDMA scheme were allocated to maximize the throughput in the downlink LiFi networks. Ling \emph{et al.} \cite{ling2018efficient}  first  showed that the problem of allocating the DC bias, the power, and the subcarriers is a coupled problem, and then  proposed several algorithms, to compromise between the performance and complexity, starting by proposing an algorithm for allocating the DC  bias only, two algorithms for allocating the power  and subcarrier jointly, and finally two algorithms  to jointly optimize the DC bias, power, and subcarrier. In \cite{wang2017ofdma}, unlike  \cite{ling2018efficient} which considered the subcarriers, authors  focused on allocating the time-frequency blocks to increase the flexibility in resource allocation. When allocating the subcarriers, channel responses should be considered, taking into account the fact that the channel quality in the low frequencies is better than that in channels at high frequencies \cite{ chen2016downlink}. Because the channel quality depends on the frequency, a careful allocation of the subcarriers  (taking the channel into account) leads to a more efficient resource allocation in OFDMA more than that in TDMA.    




%
%
%
%
%
%
%
%
%
%
%
%
%
%
%

\begin{table}[!htbp]
\centering
\label{table_LB}  
\caption{Proposed techniques to alleviate the limitations associated with VLC networks}
\begin{tabular}{| p{.08\textwidth} | p{.1\textwidth} | p{.2\textwidth}  |}

\hline
  Issue &   Solution in hybrid VLC/RF & Solution in standalone VLC \\
 \hline 

 Small coverage &   Associate uncovered users to RF network & \begin{itemize} \item CoMP \item FoV alignment \item MIMO  \item efficient APA \end{itemize}    \\
\hline
Blockages &  Associate the blocked user to RF network & \begin{itemize} \item Efficient APA \item Serve each user with multiple APs  \end{itemize} \\
\hline

Handover & Associate the unfixed users to RF network & \begin{itemize} \item Merge VLC APs to be one cell \item distribute the APs based on the anticipated location of the user \end{itemize}  \\
\hline

Interference & Associate the edge-users to RF network & \begin{itemize} \item SDMA \item Frequency reuse \item Fractional frequency or time reuse \item Joint transmission \item APs arrangement \item User-centric network design \item Efficient resource and power allocation \end{itemize} \\
\hline

Limited LEDs bandwidth & Equipping the users with multi-homing capability to gather information from RF and VLC simultaneously &  \begin{itemize} \item Efficient LED design\item extensive frequency reuse \item joint transmission \item Densify the APS and apply the user-centric design \item employ NOMA \item efficient resource allocation  \end{itemize}\\
\hline

\end{tabular}

 \end{table}

\subsection{Coordination between VLC APs}

In this section, we review the references that utilized APs cooperation techniques to improve the VLC networks. The APs in VLC networks can work together to beamform the transmitted signals, remove or mitigate interference, improve the space diversity gain, increase  coverage, decrease the handover overhead, and decrease the received SNR fluctuations.  A coordinated multi-point (CoMP) transmission technique can be implemented by connecting multiple APs through backbone networks so that they can cooperate to design their transmitted signals. Therefore, the joint transmission (JT) can be implemented between the coordinated transmitters to form one cell.

Li \emph{et al.} \cite{coop} studied how the APs should cooperate to mitigate the interference with balancing the load. 
For managing interference in the $N$ APs and $N$ users system model, the APs in the proposed system in \cite{coalition}, were designed to organize themselves into a cooperative coalition based on the game theory coalition formation. In \cite{chen2013joint}, authors adopted the joint transmission scheme to alleviate the effect of the co-channel interference and to improve the system throughput and the quality of the received signal. In addition to  the co-channel interference, the impact of blockages on users can be mitigated using  the CoMP joint transmission scheme \cite{chen2017performance}. Authors of \cite{ chen2017performance}  proposed an approach that assigns multiple transmitters to each user, with proportional fairness.  Serving a user by multiple LEDs transmitters significantly mitigates the rate of blockages and the handover overhead. 

To decrease the backbone traffic and decrease the amount of the exchanged information, authors in \cite{kashef2015coordinated} coordinated the various transmitters to control interference by either partitioning the resources among transmitters, or by controlling the transmitted power.    Partitioning the resources between transmitters decreases the spectral efficiency significantly, even though the inter-cell interference is eliminated \cite{kashef2015coordinated}. Hence, in \cite{ffr1l} and \cite{chen2015fractional}, fractional frequency reuse (FFR) was used to trade-off the spectral efficiency for the inter-cell interference, whereas Sun \emph{et al.} \cite{sun2017efficient} designed the signal for VLC system to trade-off between the interference and the spectral efficiency by imposing time superposition reuse in two neighboring cells, then they proposed an optimal power allocation strategy for this signal design approach.   Ma \emph{et al.} \cite{ma2018coordinated} exploited the spatial domain and coordinated the transmission to mitigate interference in a multi-cell MU-MISO VLC system, by considering the backbone limited capacity.  

Because only the non-negative real-valued signals can be transmitted in VLC systems, precoding techniques proposed in literature to CoMP VLC networks are different from these investigated in RF networks. Zero-forcing and dirty parity coding were investigated and compared in multi-user multiple-input-single-output (MU-MISO) VLC systems in \cite {yu2013multi} for maximizing the SINR, whereas  zero-forcing-based precoding scheme was proposed in \cite{ma2013robust, li2015multiuser, ma2015coordinated} for minimizing the mean square error. Authors in  \cite{shen2016rate} also used the zero-forcing a precoding approach for maximizing the achievable data rate, whereas authors in \cite{pham2017multi}  proposed a generalized-inverse-based zero-forcing scheme to maximize the max-min fairness and system sum rate. 

For multi-user-multiple-input-multiple-output (MU-MIMO) VLC systems, in which users are equipped with multiple PDs, the block diagonalization approach was proposed in \cite{hong2013performance} to remove interference. Pham \emph{et al.} \cite{pham2015sum} used the same precoding approach when considering the non-negativity constraint on the input signal. The Tomlinson-Harashima precoding approach  was proposed by Chen \emph{et al.} in \cite{chen2014performance}; the authors showed it outperforms the block diagonalization approach in terms of BER.  A robust linear precoding and receiver design for maximizing the minimum SINR was proposed in \cite{sifaou2017robust}. Authors in \cite{marshoud2015mu} showed that the dirty paper coding performed better than the linear precoding approaches when the users' CSI are known, whereas the linear precoding approaches are better when only an imperfect users' CSI is available.  To mitigate the effects of  the indoor VLC channel correlation, authors in  \cite{wang2015multiuser} calculated a precoding matrix for each subcarrier in a MIMO-MU-OFDM VLC system by exploiting the phase differences, after transforming them to a frequency domain of  different links. The precoding matrix was designed to eliminate the inter-user interference.  Cai \emph{et al.} \cite{cai2016photodetector} proposed algorithms of PD selection in  imaging receivers to mitigate the channel correlation and decrease the BER in a MU-MIMO-OFDM VLC system. By exploiting the knowledge of the transmitted symbols, authors of \cite{marshoud2018optical} proposed an adaptive precoding scheme to only eliminate the destructive interference and correlate the constructive interference. Designing the precoding matrix to correlate the constructive interference provides a significant improvement, in terms of BER, compared to the zero-forcing precoding approach \cite{marshoud2018optical}.  

Space division multiple access (SDMA) has been proposed in VLC networks to mitigate effects of  interference and to improve the spectral efficiency \cite{kim2014visible,yin2015low,chen2015space, chen2017space,erouglu2018multi}. In the SDMA scheme, multiple LEDs are designed to generate spatially separated beams that are directed to various users. Kin and Lee  \cite{kim2014visible} showed experimentally that the SDMA efficiently can improve the amplitude of the received signal. In \cite{yin2015low}, authors proposed a low complexity algorithm (compared to the exhaustive search algorithm) named 'random pairing algorithm' for grouping the users into multiple SDMA groups in order to obtain a better area spectral efficiency under users' fairness constraints. Each user group is served by multiple coordinating APs when applying the zero-forcing precoding method to eliminate the inter-cell interference. In \cite{chen2015space} and \cite{chen2017space},  authors proposed the use of  angle diversity transmitters  to increase the bandwidth and mitigate the interference. The authors estimated the performance of a SDMA-VLC system by deriving the analytical upper and lower bounds of the average spectral efficiency. When the number of LEDs is much larger than the number of users, where multiple LEDs can serve one user, and using the angle diversity transmitters proposed in \cite{ chen2017space}, authors in \cite{erouglu2018multi} addressed the problem of properly assigning multiple LEDs for each user. A power allocation algorithm was also proposed to improve the sum rate and the system's fairness. 


In a very dense VLC networks, especially when the number of users is much lower than the number of APs, a user-centric (UC) design is the most appropriate approach for cell formation in VLC networks.    In \cite{paving},  Zhang \emph{et al.} investigated the user centric design
for VLC, for which the cells' structures do not have a specific shape. They first clustered the users, then associated the APs to the grouped users. In \cite{user_hanzo}, Li \emph{et al.} extended the work to improve the fairness among users by proposing algorithms aimed at  scheduling users and maximizing the sum utility of the system. In \cite{hanzo_haas}, in addition to forming the cells and associating the APs, they allocated the powers to the clustered users aiming at maximizing the EE of the distributed cells. In \cite{vdo_hanzo}, authors used these techniques of cell formation and power allocation to design energy-efficient scalable video streaming with considering an adaptive modulation mode assignment. The common clustering approach used in \cite{user_hanzo,hanzo_haas,vdo_hanzo} is the edge distance clustering.
After the users are clustered, the APs are assigned to the clustered users, using  an anchoring AP association approach. Finally, the power at the APs is allocated to the associated users to maximize the EE. 

 Authors of \cite{TWC} and \cite{ICC} showed that the procedures user clustering, AP association, and power allocation are joint problems, when the EE maximization is the target. Hence, they proposed a novel user clustering method to maximize the separation between clusters and help in reduce the inter-cell interference. They then proposed a joint power allocation and AP association to maximize the EE. Inside the formed cells, two common transmission schemes were used in these  papers that adopt the user-centric design: either use the combining or the vectored transmission schemes. 
 


%

\section{NOMA in VLC}
In this section, we introduce NOMA, a new technology nominated  for the fifth generation (5G) wireless networks aimed at increase the throughput, decrease the latency, and improve the fairness and connectivity. The rational behind  NOMA is the use of a single resource component by multiple users, whether this component is a sub-carrier, a time slot, or a spreading code. With this basic concept, different types of NOMAs, such as the power domain NOMA (PD-NOMA), pattern division multiple access (PDMA), sparse code multiple access (SCMA), were presented as good candidates for the 5G multiple access technique. More details on NOMA in  traditional RF networks are provided in \cite{ding2017survey, liu2017nonorthogonal}.

In VLC networks, researchers are interested only in power domain NOMA (PD-NOMA). The goal of PD-NOMA is to set different power levels for  different users. For instance, for two users served by the same base station (BS), and using the same OFDM\ subcarriers, the BS assigns a high power to the user with poor channel and a low power for the user with a better channel. In other words, assuming that $h_1>h_2$, where $h_i$ is the  channel of the $i^{th}$ user, the BS transmits the signal of User 2 with higher power. User 2 decodes the received signal and treats User 1's signal as noise, whereas User 1  first decodes the signal of  User 2, and then removes it from the received signal, after that it decodes his own signal. To generalize this idea, we assume that we have $N$ users served by the same BS, and first categorize them based on their channel gains as $h_1\leq h_2\leq ....\leq h_N$. When using the NOMA\ technique, the BS transmits the signal of all users using same carrier, and the received signal, at the $k^{th}$ user, can be expressed as follows:

\begin{figure}[t]
\centering
\includegraphics[width=3.5in]{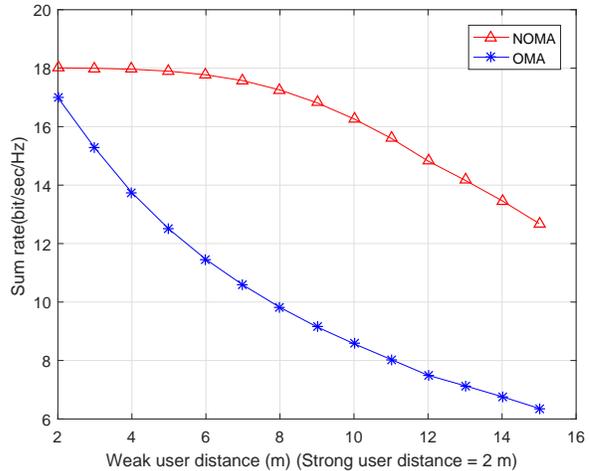}
\caption{The impact of increasing the distance of the weak user with achieving the fairness constraint that both users receive equal data rate, when FoV = 40, incidence angle = 0, and irradiance angle = 0.}
\label{NOMA_D}
\end{figure}

\begin{figure}[t]
\centering
\includegraphics[width=3.5in]{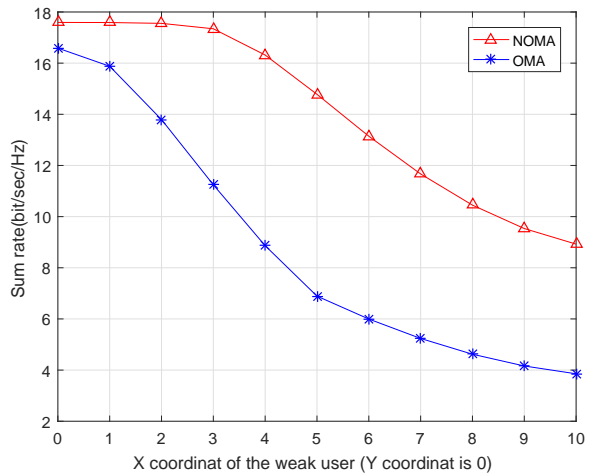}
\caption{Shifting the weak user in the X coordinate where the incidence and irradiance angles changes accordingly, the strong user located at (0,0) coordinate, FoV = 40.}
\label{NOMA_H}
\end{figure}

\begin{equation}
y_k=h_k\sum_{j=1}^N\alpha_j\sqrt{P}s_j+n_k,
\end{equation}
where $\alpha_j$ is the power coefficient of the user $j$, $s_j$ is the information signal of the user $j$, and $n_k$ is the additive white Gaussian noise. According to NOMA, users with a lower channel gain will have a higher power, meaning that $\alpha_1\geq\alpha_2 \geq ....\geq \alpha_N$. Then, the successive interference cancellation is implemented to decode the signals received by the users. In other words, User $N$ must decode all the signals of all users to have his own signal, and User $N-i$ has to decode $N-i$\ signals to obtain its intended signal. It is clear that, as the number of users increases, the complexity of decoding the signal is increased. In addition, the residual interference coming from inaccurate channel estimation  increases with the number of users.

Implementing the NOMA in VLC networks requires considering the unique properties of VLC  networks such as the limited bandwidth of LEDs, the maximum transmit power that is restricted by the illumination requirements, the blockages that make the channel between the transmitter and receiver close to zero, and the dramatic deterioration in the channel, as the  distance increases. In addition,  the channel value can be controlled by changing the FoV of the  receivers or the semi-angles of the transmitters  (if they are tunable), and these two factors can be selected to improve the performance of NOMA-VLC networks. Because the PD-NOMA scheme is  based on successive interference cancellation (SIC), NOMA-VLC networks require all users' CSIs to be available, which is the case in VLC. It was also shown that the NOMA scheme performance is enhanced as the SNR increases \cite{ding2014performance}, which is the case of VLC link. These features offered by NOMA-VLC networks led many researchers to investigate these networks and find out how the NOMA outperforms OMA schemes in VLC systems.  Figures \ref{NOMA_D}
 and \ref{NOMA_H} present simulation of how the NOMA outperforms OMA in VLC networks, for two users and one AP VLC system, when only the distance of the weak user increases (Fig. \ref{NOMA_D}) and when the distance, incidence and irradiance angles are changed (Fig. \ref{NOMA_H}). 'Strong' and 'weak' users  mean the user with the best channel and the user with the worst channel, respectively. 
 In \cite{huang2017symbol}, Huang \emph{et al.} proposed a mathematical expression of the symbol error rate analysis. In \cite{kizilirmak2015non} and \cite{lin2017noma}, authors showed the superiority of NOMA over OFDMA, in VLC systems, with respect to sum rate and BER performance, respectively. In order to allocate the power that maximizes the sum rate, authors of \cite{ shen2017optimal} optimized the  NOMA-VLC downlink for a two-user system, with satisfying certain QoS constraints. They also provided a semi-closed form to the optimal power allocation. 

In \cite{ yin2016performance, yin2015performance, yang2017fair, yapici2018non, marshoud2016non }, authors  evaluated the performance of the NOMA-VLC for one VLC AP and multiple users. In \cite{ yin2016performance, yin2015performance}, authors presented distribution functions for the uniformly distributed users, then evaluated the NOMA-VLC system by comparing it to OMA-VLC system in two case scenarios: 1) when each user has a data rate target, and 2) when the data rates of all users are assigned opportunistically according to their channels.  By considering the proportional fairness \cite{kelly1997charging}, authors of \cite{yang2017fair} showed that the formulated problem was of non-convex type, but could be converted to a convex problem that could be solved using a dual decomposition method.  Authors of \cite{ yapici2018non} evaluated and compared the NOMA and OMA schemes, when the users change their locations and their vertical orientation.  Instead of reporting the full CSI that increases the computational complexity, they used  limited-feedback schemes to categorize users based on their mean vertical angle and mean distance, and this might be most appropriate to simplify the implementation. 

For multiple APs, authors of \cite{ marshoud2016non } studied NOMA-VLC networks when the network consisting of two VLC APs and three users.  They proposed a gain ration power allocation (GRPA) approach to allocate power to the various users, and compare it with the static power allocation approach; assuming that the users' movement is assumed to follow random walk model. In the GRPA approach, the power for the user $k$ is assigned to be $P_k=(\frac{h_1}{h_k})^kP_{k-1}$. Assuming the users' FoV and the transmission angels of LEDs are tunable provides a potential to improve the performance significantly. 
For multi-cell VLC networks, under the assumption that the frequency reuse FR = 2, the users in \cite{zhang2017user} were grouped based on the received interference. If any user suffered from interference, they were given a special resource blocks, and NOMA was implemented for the remaining users.  For the users sharing the same resource block, the authors formulated  optimization problems to allocate the power, under QoS constraints, and provided solutions to the formulated problems. 

NOMA has been also used in MIMO VLC systems\cite{lin2017experimental,chen2017performance2}. In \cite{lin2017experimental}, authors  experimentally investigated a system with single carrier mode of transmission, using the frequency domain SIC, but without considering the power allocation problem. In \cite{chen2017performance2},  Chen \emph{et al.} proposed a power allocation algorithm (called normalized gain difference power allocation (NGDP) approach) aimed at reducing the complexity and increasing the efficiency of 2x2 MIMO-NOMA-VLC systems with multiple users. In their study, they classified the users using the sum of the channel gains for each user, with respect to each LED.  In the proposed power allocation method, the power allocated to the user $k$ and user $k+1$ in the $i$th LED are related:
$ P_{i,k}=(\frac{h_{1i,1}+h_{2i,1}-h_{1i,k=1}-h_{2i,k+1}}{h_{1i,1}+h_{2i,1}})^kP_{2i,k+1}$, where  $h_{ji,k}$ is the channel between the $i$th LED and the $j$th PD of the $k$th user.  

 For the uplink VLC systems, authors of \cite{guan2016non, guan2017joint} introduced a phase pre-distortion method to decrease the uplink error rate performance in NOMA-VLC systems. Since in  \cite{guan2016non}, authors used the SIC to decode the signals, whereas in  \cite{guan2017joint}, a joint detection method was used to improve the system performance. 

\section{Energy harvesting in VLC systems}
\label{EHS}


Much attention has recently been paid to energy-harvesting techniques at user-equipment devices, either from exploiting the surrounding environment, or by transferring wireless power. Energy harvesting is the capability of converting the radio frequency (RF) signals or light intensity into electrical voltage/current. With the advent of the era of the IoT, the demand for transferring the power and enabling IoT devices to harvest  energy using  light or RF transmission is increasing, especially in indoor applications where smart buildings, health monitoring, and sensors devices applications become abundant.
\begin{figure*} 
    \centering
  \subfloat[FoV=30$^o$]{
       \includegraphics[width=0.49\linewidth]{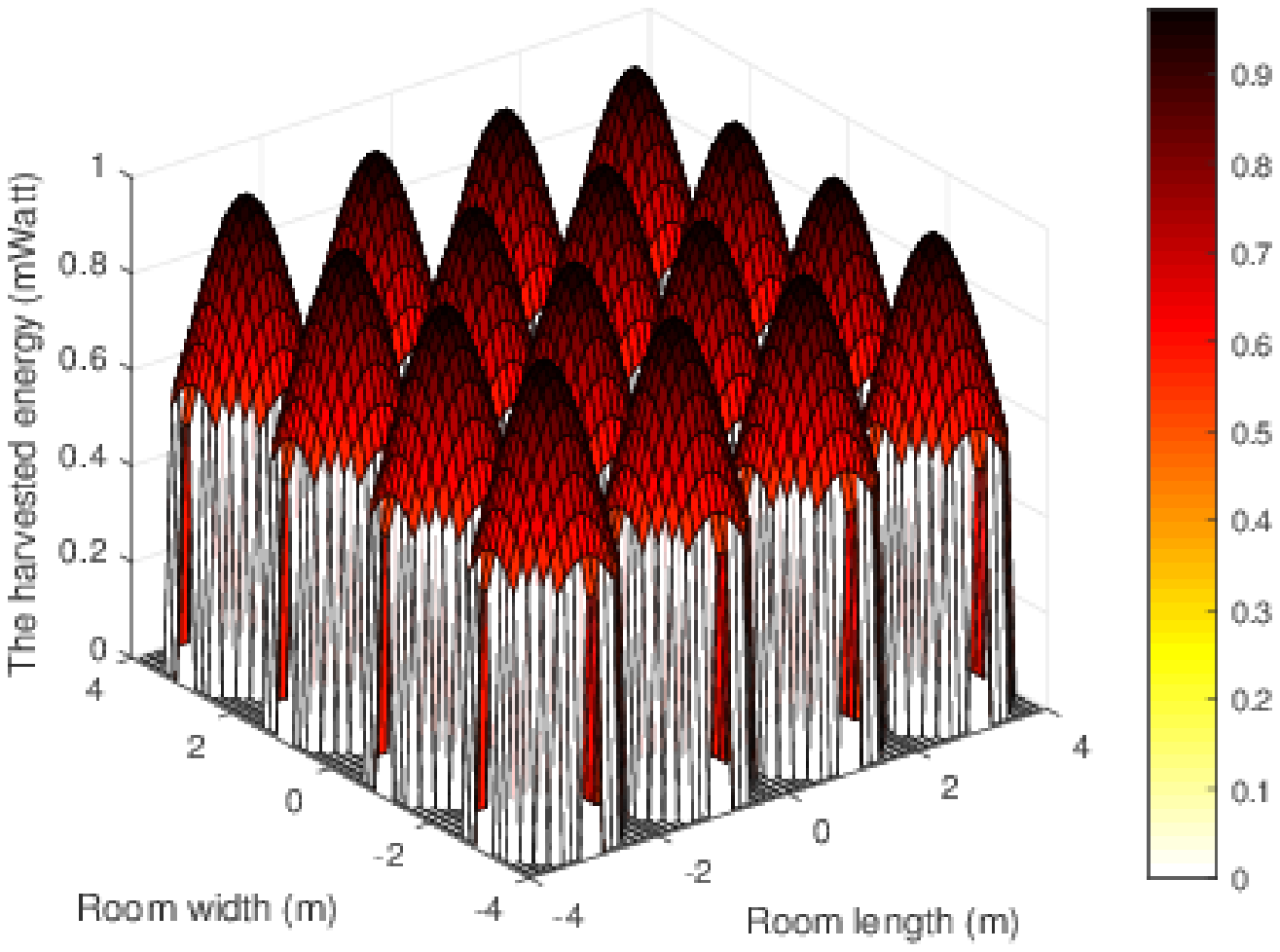}}
    \label{1aa}\hfill
  \subfloat[FoV=40$^o$]{
        \includegraphics[width=0.49\linewidth]{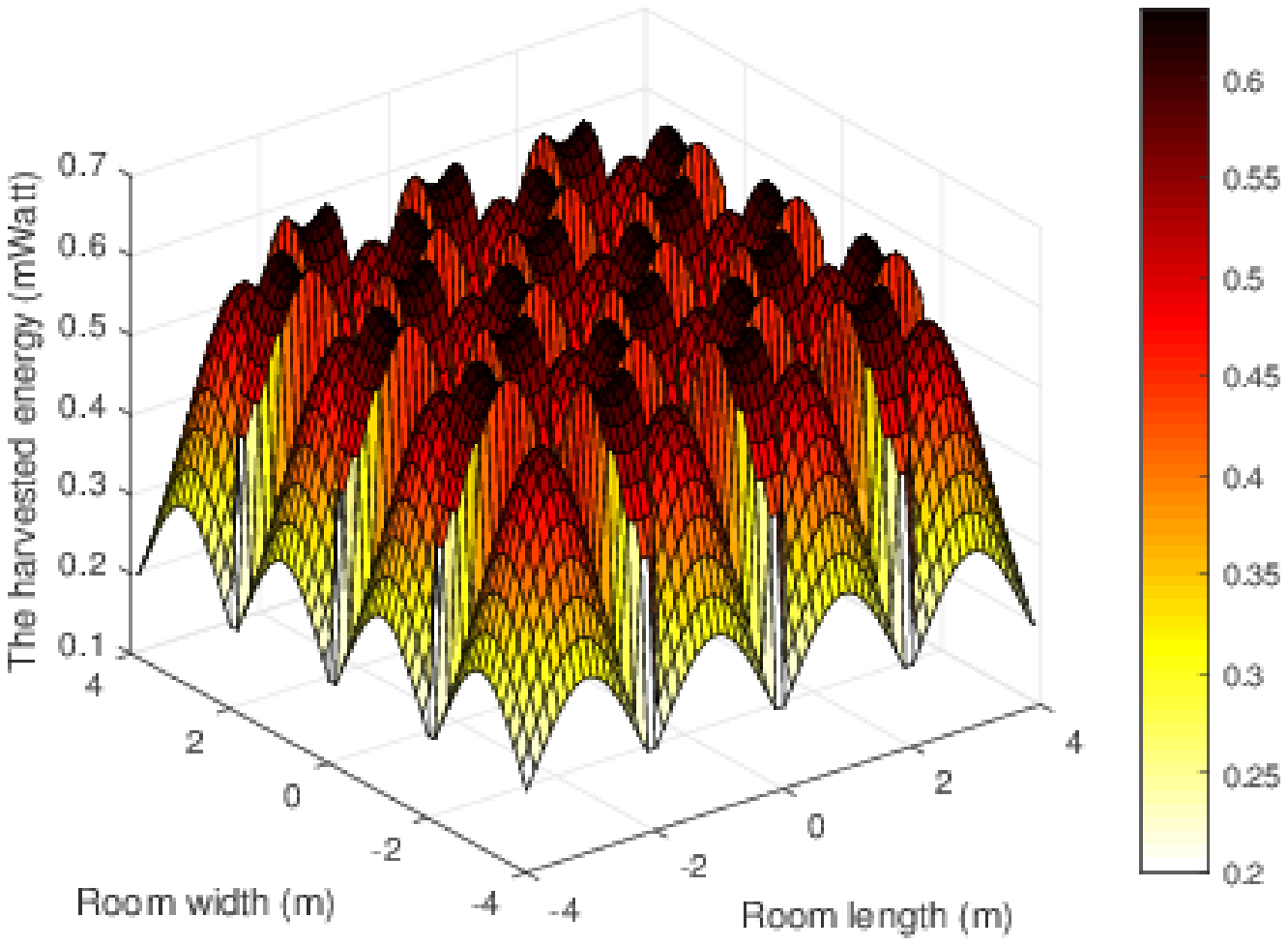}}
    \label{1bb}\\
  \subfloat[FoV=50$^o$]{
        \includegraphics[width=0.49\linewidth]{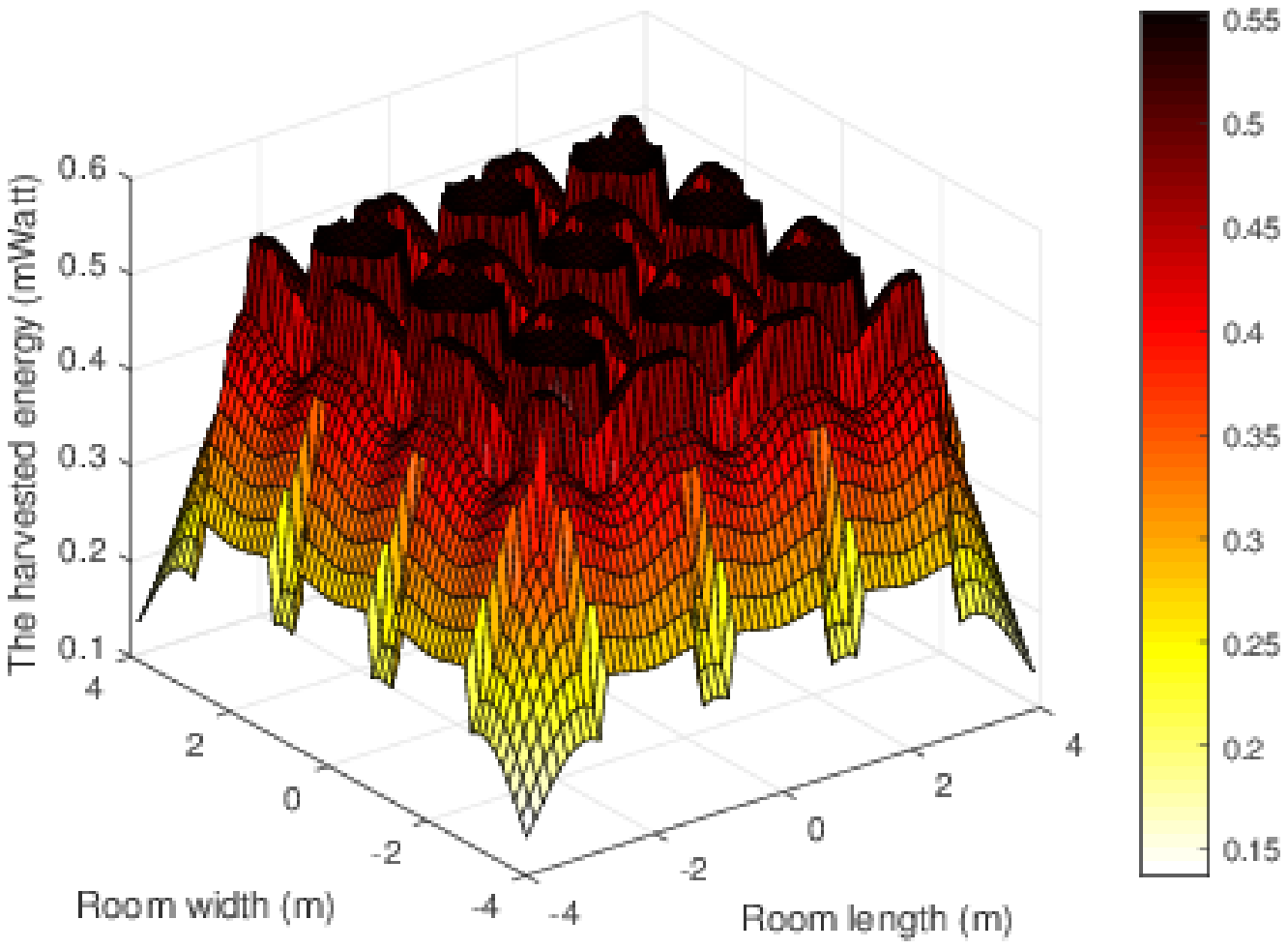}}
    \label{1cc}\hfill
  \subfloat[FoV=60$^o$]{
        \includegraphics[width=0.49\linewidth]{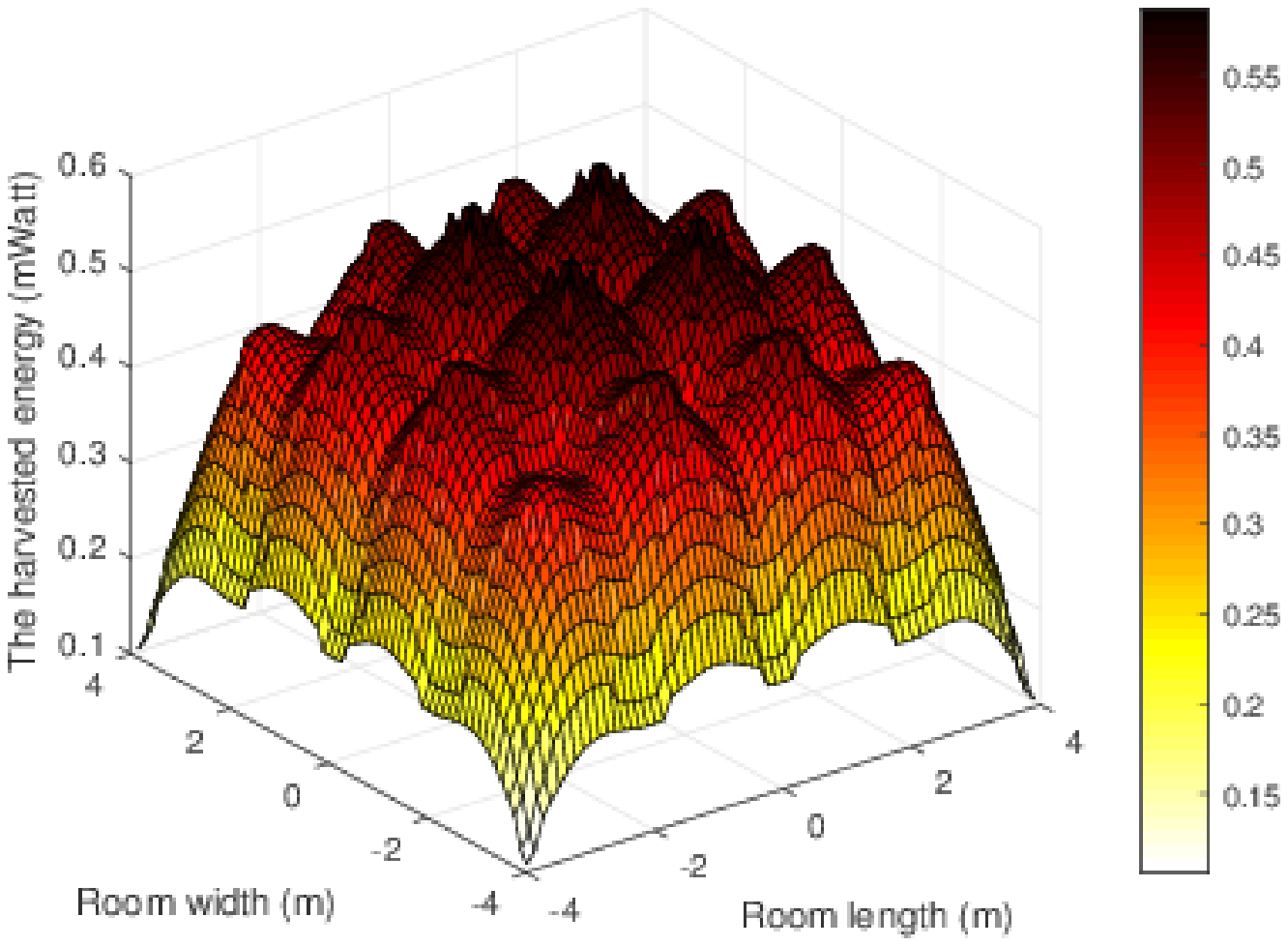}}
     \label{1dd} 
  \caption{The distribution of the possible harvested energy over the area, with diferent user's FoV, where the system model consists of 16 VLC APs.}
  \label{fig1} 
\end{figure*}
 Extensive work has been done to design, evaluate, and optimize  simultaneous wireless information and power transfer in RF networks \cite{perera2017simultaneous,lu2015wireless}. Work on harvesting the energy and transmitting the information using the light is scarce, as it is still in its early stage. 

Different from harvesting the energy in the RF networks, the energy can be harvested using the DC component that is transmitted along with the information signal to ensure the non-negativity of signals. This DC component can be easily separated from the modulated signal using capacitor and goes to the energy harvesting circuit. Since the recent solar cell panels can provide more than 40\% conversion efficiency \cite{king2012solar}, a new trend has emerged in the research community for using solar cells at the receivers to detect information signal and harvest energy. 

Investigating the harvesting of energy in VLC systems has been a timely topic of interest \cite{solar,fakidis2016indoor, dual1,dual2,sum_rate, amr,on3, simul2}. A few papers   recently published  proposed to investigate systems that use the light to jointly transfer  power, meet illumination requirements, and transmit data. Authors of \cite{solar}  experimentally harvested the solar energy with mobile phone by equipping it with a commercial solar panel in an indoor environment. They showed that the devices directly exposed to the  indoor light could be charged to a satisfactory level. Fig \ref{fig1} shows the amount of the possible energy, which can be harvested by a user, over the surface of an $8 \times 8$ room with 16 VLC APs. As observed Fig.\ref{fig1}, the user's FoV has an impact on the harvested energy. 

Authors of \cite{fakidis2016indoor} investigated the concept of indoor optical wireless power transfer to solar cells during darkness hours. By using laser diodes and a solar panel, they measured the power efficiency and  showed an improvement over the inductive power transfer systems, of approximately 2.7 times. By using 42 laser diodes, they claimed to deliver 7.2 W of optical power to a solar panel $30$ m distant from the diodes. Authors of \cite{carvalho2014feasibility} studied how much artificial indoor light could deliver an amount of energy, using different types of receiving cells.  

In \cite{dual1} and \cite{dual2}, a dual-hop hybrid VLC/RF communication system was studied as a means to reach out to the out-of-the-coverage user. The authors showed that visible light could be used, in the first hop, to transfer both data information and  energy to the relay. The relay, could then forward the data to the destination, using the harvested energy. In \cite{sum_rate,amr}, authors maximized the sum-rate utility of a VLC system consisting of one AP and $K$ users, subject to individual QoS constraints. Li \emph{et al.} \cite{sum_rate} assumed that a user $k$ van receive the information in their assigned time slot, and the power within the time slots assigned to other users.  In \cite{amr}, on the other hand, Abdelhady \emph{e al.} proposed solving the problem by allocating the optical intensity and time slots, using an  upper bound on the individual required harvested energy.
Authors in \cite{on3} characterized the outage performance of a hybrid VLC-RF system, where the visible light is used for the downlink to transfer the energy and data to the users, who then use the harvested energy to transmit a RF signal in the uplink. 

All the studies mentioned above use the alternating current (AC) component for harvesting the energy, where the DC component of the transmitted light is fixed and readily used to harvest energy \cite{wang2015design}. In \cite{wang2015design}, authors designed an optical wireless receiver using a solar panel and enabled it to receive information and harvest energy simultaneously. Because the received current signal contains both DC-current and the AC-current components, authors in \cite{wang2015design} suggested to attenuate the AC current, using an inductor to remove the ripples from the DC-current that is forwarded to energy harvesting branch, and to block the DC-current, using a capacitor, to obtain only an AC-current in the communication circuit. Sandalidis \emph{et al.} \cite{ sandalidis2017illumination} investigated the three functions of the LED lamp, i.e. the illumination, communication, and the energy harvesting, on a system consisting of a desk LED close to the receiver equipped with a solar cell. The authors divided the received optical power  between the information signal and the harvested energy, using a splitter. 

Diamantoulakis \emph{et al.} \cite{simul2} studied the lightwave information and power transfer for a system consisting of one transmitter and one receiver. They provided the two following protocols to maximize the harvested energy at the receiver, under data rate constraint: 1) Splitting the time into two portions, one dedicated to maximize the user's SINR, and the second assigned purely to maximize the harvested energy by maximizing the DC component, 2) Optimization of the DC bias, in phase 1, to maximize the harvested energy under QoS constraints, with phase 2  assigned only to harvesting the energy. However, optimizing the DC bias for the whole time is more general, and there is no need to split the time between harvesting the energy and transmitting the data. In addition, the formulated problem would be more challenging if there were multiple receivers, since the fairness, in terms of data rate and harvested energy, is required. Hence, authors in \cite{globecom2} studied the simultaneous information and power transfer in MISO multi-user VLC system, where some users are interested only in gathering information and the others users interested in harvesting energy. The problem of maximizing the total harvested energy under QoS constraints was formulated as a non-convex problem, and hence, a special iterative algorithm was used to solve the formulated problem.

\section{Securing VLC systems}
Traditionally, the mission of the physical layer researchers and designers is to provide a reliable signal transmission to the relevant receivers, while the mission of securing and protecting the transmitted information is usually assigned to the upper layers of the networks \cite{massey1988introduction}. However, with the increasing  demand for  high data rate and the spreading of the broadcast-nature networks, researchers have found with new mechanisms to build secure communication networks, using the physical layer. 

The physical layer security (PLS) deals with how to exploit the randomness of noise, channels, and different resources, such as multi-antenna and cooperative nodes, to minimize the information that can be extracted by the eavesdropper \cite{zhou2016physical,obeed2018efficient}. 
The secrecy capacity was introduced by Wyner \cite{wyner1975wire} as a metric to measure the security performance; it was defined as the highest information rate that can be acquired at the legitimate receiver, with having the eavesdropper completely unaware of the transmitted information. To have a precise quantification of the security, the eavesdroppers are assumed to have unlimited knowledge of the network parameters and have a sufficient computational capability.  

In this section, we  review the work that has been performed on the PLS, in VLC networks. For the PLS in RF networks, readers can refer to \cite{mukherjee2014principles} and \cite{chen2017survey}.

As mentioned in the Introduction Section VLC networks are more secure than RF networks and less susceptible to signal interception because of the small coverage provided by LEDs, and because they work properly only in the presence of the LoS components. However, the security in VLC networks is still a problematic issue, specifically when transmitted information can be accessed by multiple users, for instance in public areas, meeting rooms, laboratories, and libraries. This means  that potential eavesdroppers may be able to gather confidential messages \cite{mostafa2015enhancing}.  

Many papers reported in the literature address the PLS in VLC networks, and different techniques were proposed to evaluate and improve the secure communications. The proposed techniques generally depend on different network parameters, such as  the availability of the eavesdroppers' CSI, the number of LEDs equipped with transmitters,  the number of the legitimate users and eavesdroppers. These techniques use a zero-forcing precoding approach that eliminates the transmitted information to eavesdroppers; they may use artificial noise or jamming aimed at confusing eavesdroppers, they build a protected zones using angle diversity transmitters, optimizing the input distribution,  characterizing the security in VLC systems using the stochastic geometry. The artificial noise and signal modification are possible in real-world VLC applications, but the MAC layer cannot provide sufficient protection against eavesdropping \cite{blinowski2016practical}. To our knowledge, to this day, a closed-form expression of secrecy capacity in the VLC networks has not been derived. All the studies reported in the literature  derive a lower and upper bounds for the secrecy capacity and for the secrecy outage probability.     

Unlike  RF channels, the optical intensity must be considered for  illumination requirements. Hence, in VLC channels, the optical intensity is a constraint that is directly proportional to the electrical signal amplitude, not to the squared signal as in  RF channels. 

In \cite{mostafa2014physical}, Ayman Mostafa and Lutz Lampe  started investigating the PLS in MISO-VLC channel with one eavesdropper and one authorized user. When the CSI of the eavesdropper is available, the zero-forcing precoding approach is applied, but if the CSI is unknown, the transmitter divides its own optical power into two portions, a one used for the information-bearing signal and the other used for emitting jamming signals to confuse eavesdroppers without affecting legitimate receiver. It is important to note that the zero-forcing preceding approach can be implemented when the number of LEDs transmitters is larger than the number of eavesdroppers. Similarly, the jamming signals can be eliminated, at the legitimate users, when the number of users is lower than the number of LEDs jammers. 

Instead of dividing the power between information and jamming signals, Ayman Mostafa and Lutz Lampe  \cite{mostafa2014securing} assigned one LED for transmitting data and assigned all the others to transmit  jamming signals. To eliminate the jamming signal at the legitimate receiver, different LED jammers coordinate their transmitted signals to be eliminated at the legitimate receiver. In \cite{chow2015secure}, authors devoted some LEDs to transmit the information and keeping the others for emitting an intrusion signal and shape a protected zone where potential eavesdroppers receive a degraded SNR. 

In \cite{mostafa2015physical2}, authors presented a formulation for the upper and lower bounds of the secrecy rate, under amplitude constraint for the SISO-VLC system; they then generalized the capacity bounds to the MISO systems. They also solved two optimization problems to find the optimal weighting vector at the light source transmitters, when CSI of the eavesdropper is known and when the transmitter has a limited information about the location of the eavesdropper. For amplitude-constrained wiretap channels, the secrecy rate maximization problem was formulated as a non-convex optimization problem when the CSI of the eavesdropper is known \cite{mostafa2016optimal}.  The authors transformed the nonconvex problem into a solvable quasiconvex line search algorithm that provides a slightly better performance than the simpler zero-forcing algorithm proposed in \cite{mostafa2015physical2}. When the CSI of the legitimate user is limited, and the CSI of the eavesdropper is not available, they estimated the channel of the eavesdropper by deriving the uncertainty sets that reflect the inaccurate knowledge of the eavesdropper location, orientation, LEDs half angle, and non-LoS components. 
In \cite{shen2016secrecy}, a joint beamforming and jamming approach was proposed to improve the PLS for one legitimate user with multiple eavesdroppers in the MISO-VLC system. Two vectors were optimized jointly: the beamforming vector that is multiplied by the legitimate information, and the jamming precoding vector that is multiplied by the jamming signal, in order to maximize the SNR of the legitimate user, under SNR constraints for both perfect and imperfect  CSIs at the transmitter. 

In \cite{pham2016secrecy} and \cite{pham2017secrecy }, V. Pham and T. Pham investigated ways of  transmitting  confidential messages to  two different legitimate users,  in a MISO-VLC  system, while keeping the messages confidential from each other (as well as the eavesdropper). The zero-forcing precoding approach was applied to guarantee the confidentiality among users.  When the transmitter wants to transmit $K$ confidential messages to $K$ users, a new precoding scheme was proposed in \cite{arfaoui2017achievable2}
 to maximize the secrecy sum rate, in a multi-user MISO VLC system. This  precoding scheme is based on finding the beamforming matrix from the eigenvectors associated with the largest eigenvalues of the different $K$ MISO-VLC channels. In \cite{mostafa2015pattern}, authors studied the PLS in a system consisting of massive low-intensity LEDs, distributed uniformly in the ceiling, multiple legitimate receivers, and multiple eavesdroppers with limited CSI.  After defining the insecurity zone around the legitimate receivers, they designed the beamformer so that it can direct its main lobes towards the defined zones and minimize the information rate outside the defined insecurity zone. 
 
 Because of the peak amplitude constraint on the input distribution, the Gaussian input distribution is not the optimal option. Therefore, different input distributions,  under intensity constraints, were proposed to improve the secrecy performance in VLC networks \cite{mostafa2014securing,zaid2015improved,mostafa2015pattern, arfaoui2016secrecy,arfaoui2016input}. Specifically, authors of \cite{mostafa2014securing } used the uniform input distribution.  Authors of  \cite{zaid2015improved}  improved the work of \cite{mostafa2015pattern} by using the truncated Gaussian inputs instead of the uniform input distributions for both the data and the artificial noise signal. In \cite{arfaoui2016secrecy}, authors used the a truncated generalized normal distribution for the MISO-VLC Gaussian channel. It was shown that the truncated generalized normal distribution input performs better than the uniform and the Gaussian  distribution input \cite{arfaoui2016secrecy}. For the same system model and same input distribution as in \cite{arfaoui2016secrecy}, Aefaoui \emph{et al.}  \cite{arfaoui2016input} found a closed-form for the optimal beamforming, for a known location of the eavesdropper. Both the truncated Gaussian distribution and the truncated generalized normal distribution provide potential to improve the secure communications by optimizing the parameters that cannot be optimized in the uniform distribution. 


When the VLC system contains multiple eavesdroppers distributed randomly in the considered area, the stochastic geometry can be used to characterize the system and to provide analytical expressions of the secrecy outage probability and of the average secrecy rate. In \cite{pan2017secureon}, authors studied the PLS in one VLC cell consisting of  group of LED lamps located in the center of the ceiling, with one receiver and multiple eavesdroppers. The eavesdroppers were randomly distributed based on the Poisson point process. They provided  closed-form analytical expressions for the secrecy outage probability and the average secrecy rate, using the stochastic geometry method under the assumption that the floor area is a circle.

In \cite{cho2017secrecy}, Cho \emph{et al.} proposed a LED selection scheme to improve the secrecy outage probability, when the eavesdroppers are randomly located in the selected area, and their CSI  are unknown to the transmitters. Similarly to what was done in \cite{pan2017secureon}, they  used the Poisson point process to model the randomness of the location of the eavesdroppers. The same authors used the proposed eavesdropper location modeling to analyze the performance of the MISO-VLC system model and proposed a beamforming solution \cite{securing}. To find the optimal beamforming vector, they formulated the following optimization problems: minimizing the average eavesdropper SNR under a given SNR constraint to the legitimate receiver, maximizing the authorized user SNR under the average eavesdropper SNR constraint, and they optimized the same problems for the rate, instead of the SNR. In all the formulated problems, closed-form expression were provided for the beamforming vector solution. Because the VLC LEDs provide a very limited coverage area, the beamforming solution for the secrecy maximization can be approximated to be a LED selection, suggesting  the selection of the AP closest to the legitimate user;  secrecy rate would then be  improved,  as  the distance between the eavesdropper and the legitimate user increased  \cite{securing}. The same authors  studied the PLS in a VLC system when multiple eavesdroppers combine their observations using maximum ratio combining (MRC) approach, to maximize their information rate and degrade the secrecy performance \cite{cho2018physical}. The authors used the stochastic geometry to anticipate the secure communication under a pre-defined eavesdroppers' density.

In \cite{yin2018physical}, Yin and Haas considered the unique properties of the VLC channel and the VLC network's layout to fully characterize the secrecy outage probability and the ergodic secrecy rate in multiuser (multiple legitimate users and multiple eavesdroppers), multi-cell VLC systems. The APs in the ceiling were modeled using a two-dimensional homogeneous Poisson point process, under the assumption that some of the LEDs are not working as  VLC APs (they provide only the illumination), whereas both users and the eavesdroppers are modeled using another independent two-dimensional homogenous Poisson point process at their plane. They investigated the secrecy outage probability and the average secrecy capacity for three scenarios: 1) the legitimate user is served by the nearest AP, 2) the legitimate user is served by  cooperating APs, and 3) the legitimate user is located in the protected zone around the AP, where the eavesdropper is not allowed to be. 

Because all the aforementioned work ignored the fact that an eavesdropper might exploit the reflected light to obtain unauthorized information, authors of \cite{liu2016new} considered the effect of the reflected path and the channel correlation in a MISO-VLC network to propose an eavesdropping-resilient framework for VLC security. Authors of \cite{classen2015spy} showed that a small gap under the door, kay holes, or the window could be  sufficient sources for eavesdropping. They also showed  that eavesdroppers could gain information from  reflected lights from the walls. Cho \emph{et al.} \cite{cho2018impact} showed how the non-LOS or the reflected light would affect the secrecy outage probability in a system model consisting of multiple LED transmitters, one legitimate user, and multiple randomly distributed eavesdroppers. They showed that the secrecy outage probability depends on the position of the legitimate user,  the design of the LED transmitters, and the location of the eavesdroppers with respect to the reflecting points. 

For MIMO-VLC systems, in \cite{le2014secured}, a MIMO system was used to establish a secure communication zone by minimizing the BER in the protected zone and maximize it everywhere else. In \cite{arfaoui2017achievable}, authors studied the PLS in a MIMO-VLC system model consisting of one transmitter equipped with multiple LEDs, one multiple-PD eavesdropper, and one multiple-PD authorized user. In both cases (the CSI of the eavesdropper is known or unknown), they derived the optimal covariance matrix and the optimal signaling scheme for achievable secrecy rate maximization, then  derived an upper bound for the proposed system secrecy capacity. Authors of \cite{chen2017physical} improved the PLS in VLC systems by applying angle diversity transmitters that are capable of transmitting data in  narrow beams, effectively minimizing the leakage of the information. By comparing  different types of optical network deployments, they concluded that the hexagonal deployment is the best in terms of secure communications, whereas the Poisson point process deployment is the worst. Wang \emph{et al.} \cite{wang2018secrecy} proposed LED pattern selection scheme to improve the secrecy performance of the generalized space-shift keying VLC systems. 

For the hybrid VLC/RF networks, authors of \cite{marzban2017beamforming} studied the PLS when the network consisted of one RF AP, one VLC AP, one legitimate receiver, and one eavesdropper. Both the eavesdropper and the receiver have the multi-homing capability (i.e. they can aggregate the information from both networks). The RF AP and the VLC AP are equipped with multiple antennas and multiple LEDs, respectively. By optimizing the beamforming vectors and the transmit powers at both the VLC and RF APs, they formulated their problem as minimization of the total consumed power under having the aggregated information rate at the eavesdropper nulled and having the information rate at the receiver above a predefined threshold. In \cite{pan2017secure} and \cite{pan2017secrecy}, authors derived the exact and the symptomatic secrecy outage probability for the uplink transmission in a system consisting of one legitimate receiver, one eavesdropper and two RF VLC APs. The downlink was implemented by the VLC AP, where the receiver and the eavesdropper harvest the energy from the light intensity. Then, during the receiver transmission of the information in the RF link with a finite energy storage, the eavesdropper tried to acquire the transmitted information. They concluded that increasing the circle area or decreasing the LED height improves the secrecy outage probability. 

In \cite{mukherjee2016secret},  Mukherjee investigated the lower and upper bounds of the secret-key capacities in SISO VLC system, and analyzed the secret-key transmission scenario in MISO VLC system. Al-Moliki \emph{et al.} proposed a security protocol that generates confidential keys dynamically from the bipolar real OFDM samples to encode each signal frame by using the cyclic prefix samples placed in the small channel impact area \cite{al2016robust,al2017secret}. The same authors upgraded their proposed protocol in \cite{ al2016robust } to combat the known-plain text attacks and chosen-plaintext attacks by applying logistic chaotic maps to the system \cite{al2017physical}.

A chaotic channel determined subcarrier shifting (CDSS) scheme with pre-equalization were presented in \cite{lu2016design} to improve the PLS in DCO-OFDM VLC system. For secure image transmitting in VLC channel, chaos scrambling schemes were proposed in \cite{ wang2017secure } for discrete-Foruire transform precoded OFDM-based, and in \cite{wang2018two} for discrete cosine transform precoded OFDM-based systems.
\begin{table}[!t]
\centering
\caption{Techniques used for PLS}
\label{table_PLS}
\begin{tabular}{|p{2cm} | p{5 cm} |}
\hline
  State & Appropriate techniques \\
 \hline 

 Full CSI is available & \begin{itemize} \item  Zero-Forcing precoding;\item design the beamforming matrix from the channel matrix eigenvectors; \item LED selection \end{itemize}   \\
 \hline
 
 Location of the eavesdropper is available &  \begin{itemize} \item build a protected zone where eavesdropper cannot be located;  \item jam a defined zone where the eavesdropper can be located;   \item estimate the eavesdroppers' channel and use the same techniques when CSI is available;  \item use the angle diversity transmitters; LED selection; APs arrangement\end{itemize} \\
 
 \hline
 
 No information available about eavesdroppers & \begin{itemize} \item characterization of the secrecy performance using the stochastic geometry tool; \item divide the power between jamming and beamforming vectors; \item devote some LEDs for jamming; \item LEDs selection; \item APs arrangement \end{itemize} \\

  \hline
\end{tabular} 

 \end{table}
 
 \section{Summary and Open Research Problems}
 In this paper, we reviewed the optimization techniques that proposed to improve the performance of VLC networks and minimize the effects of the VLC limitations. The review covered the different types of VLC networks: hybrid VLC/RF networks, VLC standalone networks with APs' coordination, NOMA-VLC networks, VLC networks that contain energy harvesting users, and VLC networks that contain eavesdroppers.  
 
 Based on the existing work in the literature, in this section, we outline different challenges and open research problems that need to be considered and investigated in the future work.   
  \subsection{NOMA-VLC Networks}
 Despite all the aforementioned work on NOMA-VLC systems, numerous challenges remain, and important topics in this area of research are still to be investigated. Below is a list of some key open problems in NOMA-VLC networks:
 
\begin{itemize}
\item Hybrid NOMA-VLC systems: the hybrid NOMA is to group the users into multiple clusters, and assign to each cluster a designated resource block, following the NOMA principle in each cluster. To our knowledge, the hybrid NOMA has not been studied yet in VLC systems.  The rational for using the a hybrid NOMA is its ability to reduce the system's complexity. Indeed, having a large number of users in the VLC system, and assigning them to the same resource block can be problematic, since the user with the best channel must decode all the signals of all the users before decoding his/her own signal, creating delays the decoding and resulting in high complexity.  Hybrid NOMA systems have been proposed in RF networks to take into account both the system performance and complexity. Therefore, we propose to study the hybrid NOMA-VLC system by finding the optimal user grouping, allocating the power to each group, the power inside each group, or grouping the users and allocating the power jointly, for sum-rate maximization purposes. This system can be extended to be a multi-cell system, in which the user-to-AP association problem also exists and should be considered. Hence, the problem would be then a two layer user grouping with power allocation. 

\item NOMA with different QoS requirements: In real life, not all users require the same amount of data rate. For example, some of them may stream videos, whereas others are texting or exploring websites. Also some receivers can be IoT devices that need  low data rates. Allocating power, in the most effective way, to users with different needs still remain a challenge, and obtaining the required data rate for each user, even when some weak users (users with poor channels) require higher data.  

\item Cooperative NOMA-VLC: cooperative NOMA has been proposed in RF networks to exploit the redundant information in NOMA systems, and to compensate the weak user  suffering from co-channel interference by increasing his data rate. The cooperative NOMA can be used among VLC users or by using relays. We are interested in the cooperation among users, where the strong users that encode the signals of the weak users can forward these coded signals to the intended users. When assuming a VLC system with two users, under the assumption that all users are equipped with a multi-homing mechanism (i.e. they can gather information from VLC and RF networks simultaneously), the  strong user first decodes the signal of the weak user, then forwards it to the weak user using WiFi or Bluetooth techniques. Weak user can then combine the RF  and VLC signals, using combining techniques.  

\item Modulation and coding for NOMA-VLC system: several papers studied the modulation and coding schemes in NOMA RF networks \cite{shieh2016gray, shieh2016simple}. As the modulation and coding in VLC networks is different (based on IM/DD), investigating the modulation and coding schemes in NOMA-VLC systems would be worthy for practical implementation.  

\item NOMA in coordinated multi-point (CoMP) VLC networks: CoMP VLC system means that multiple APs are cooperating to transmit the data for the users. The cooperation is for mitigating the inter-cell interference and enhancing the received data rate by optimizing the precoding matrix. Assume a VLC system consisting of $N$ APs and $M$ users, where $M>N$, the questions should be raised is that how the users should be sorted from the strongest user to the weakest user?, how should the users be grouped to be served by the cooperating APs?, and how should the power be allocated. Combining the two techniques NOMA and CoMP surly leads to having a significant performance improvement in VLC systems.  


\item Hybrid SDMA and NOMA: SDMA in VLC can be implemented using  angle diversity transmitters that can generate several parallel narrow light beams directed to different directions using different LEDs.  The goal of using SDMA is to mitigate inter-cell interference in VLC networks by directing the light to intended users and decreasing the overlap areas. However, some LEDs can be directed to non users, some to one user, and others to multiple users. The LEDs that are assigned or directed to serve multiple users can use the NOMA as a multiple access and to maximize the data rate. SDAM is used to mitigate or eliminate inter-cell interference, and NOMA is used to mitigate the intra-cell interference using SIC. By combining both of them, the system performance is significantly improved, in terms of data rate and system's fairness. Fig. \ref{SD_NO}
 shows a system model in which  SDMA and NOMA can coexist in VLC systems, where the NOMA can be used in LEDs that serve more than one user.  
\begin{figure}[t]
\centering
\includegraphics[width=3in]{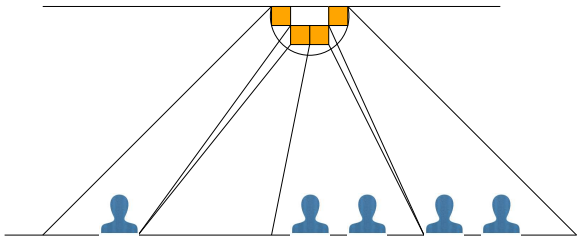}
\caption{The proposed hybrid SDMA/NOMA system.}
\label{SD_NO}
\end{figure}

\end{itemize}

\subsection{Harvesting the Energy in VLC Systems}
Despite all the aforementioned work in Section \ref{EHS}, there are several remaining challenges associated with the transfer of information and power, using  a light wave. Here are below a few key issues  that need to be investigated and optimized for obtaining the most efficient power and information transfer systems. 
\begin{itemize}
\item \emph{Simultaneous light-wave for information illumination and power transfer:} Several studies investigated  VLC systems in which both energy and information could be transferred to users. However, achieving both functions in VLC networks might violate the illumination requirements. We therefore propose to study the three functions of the light simultaneously by formulating optimization problems that allocate the DC bias, the available power, and the available resources. 
\item \emph{Joint DC-bias and resource allocation for sum rate with the presence of  energy-harvesting users}: allocating both the DC bias and the available resources at the VLC APs leads to a significant improvement of the VLC performance under simultaneous lightwave information and power transfer (SLIPT). An effective allocation of resources (to the users) provides opportunities to preserve high energy that can be harvested by users.

\item As proposed in the NOMA-VLC Section, a cooperative NOMA can be implemented in VLC systems; however, the strong user may do not want to consume some of his/her power by forwarding the signal to the weak user. We therefore suggest investigating ways for the strong user to harvest the energy  from the light intensity, in the first phase, and then use it to forward the weak user's signal. This means that the transmitter should optimize the DC bias and the information power to maximize the sum rate and guarantee acceptable fairness. 

\item \emph{Optimizing the MISO-VLC network with NOMA:} when the system consists of multiple VLC APs cooperating to transmit the information and power for multiple users, if the number of users is larger than the number of APs, the key issue that should be addressed is whether OMA or NOMA is the best system for scheduling users and harvesting the energy. As previously reported in the literature, NOMA can provide  better data rates than OMA. In other words,  NOMA can achieve the required users' data rate with a small amount of transmit power (information power), which allow the DC bias to increase, resulting in increasing the harvested energy.  

\item \emph{Placing the energy harvesting users:} suppose that a VLC system consisting of multiple information users (users interested only in gathering the information), and IoT devices that work only in uplink (like sensors) and interested only in harvesting the energy (energy harvesting (EH) users). Optimizing the positions of the EH users in order to maximize the harvested energy and to achieve the required QoS at the information users is crucial in VLC SLIPT systems. Yet, it remains challenging. Therefore, there is needed to investigate ways of implementing and simplifying this task.
\end{itemize}  

\subsection{Securing VLC Networks}
Despite the significant number of studies already performed, there are still some important issues to tackle  and still many challenges for researchers to overcome  in the future.  A few of them are highlighted below, together with potential solutions that may improve the PLS  in VLC systems:

\begin{itemize}
\item How to optimize the beamforming vector in MISO-VLS systems when an active and passive eavesdropper exists. The common approach for the active eavesdropper is the zero-forcing preacoding approach, the common approach for the passive eavesdroppers is the design of protected zones, using an artificial noise or by steering the beamforming lobes. This raises an important question: what would the appropriate method be to  improve the security, if the transmitters  know the CSI of some eavesdroppers and they do not know the CSI of the others, or have a limited information about the eavesdropper (e.g.  location only).

\item PLS in NOMA-VLC system: Several recent papers investigated the PLS in NOMA RF networks for different system models \cite{liu2017enhancing,zhang2016secrecy}. To this day, no paper has  studied the PLS in NOMA-VLC systems. Because of the unique properties of VLC systems, the PLS in NOMA-VLC systems is required to be investigated, evaluated, and optimized.

\item User-centric cell formation  based in the presence of eavesdroppers: As shown above, the user-centric cell formation is an appropriate scenario when the number of users is much smaller than the APs. Suppose that the network contains some eavesdroppers (whether their CSI are available or not), the questions raised are: 1) how should the users  be clustered? 2) how should the APs  be associated to the clustered users? 3) which  APs  should participate in communication, and which should be switched off? 4) could the switched off APs help enhance the secrecy sum-rate in emitting jamming signals?

All the above questions indicate that the joint PLS and user-centric design should be investigated and optimized together.

\end{itemize}


\bibliography{mylib}

\bibliographystyle{IEEEtran}

\begin{IEEEbiography}[{\includegraphics[width=1in,height=1.25in,clip,keepaspectratio]{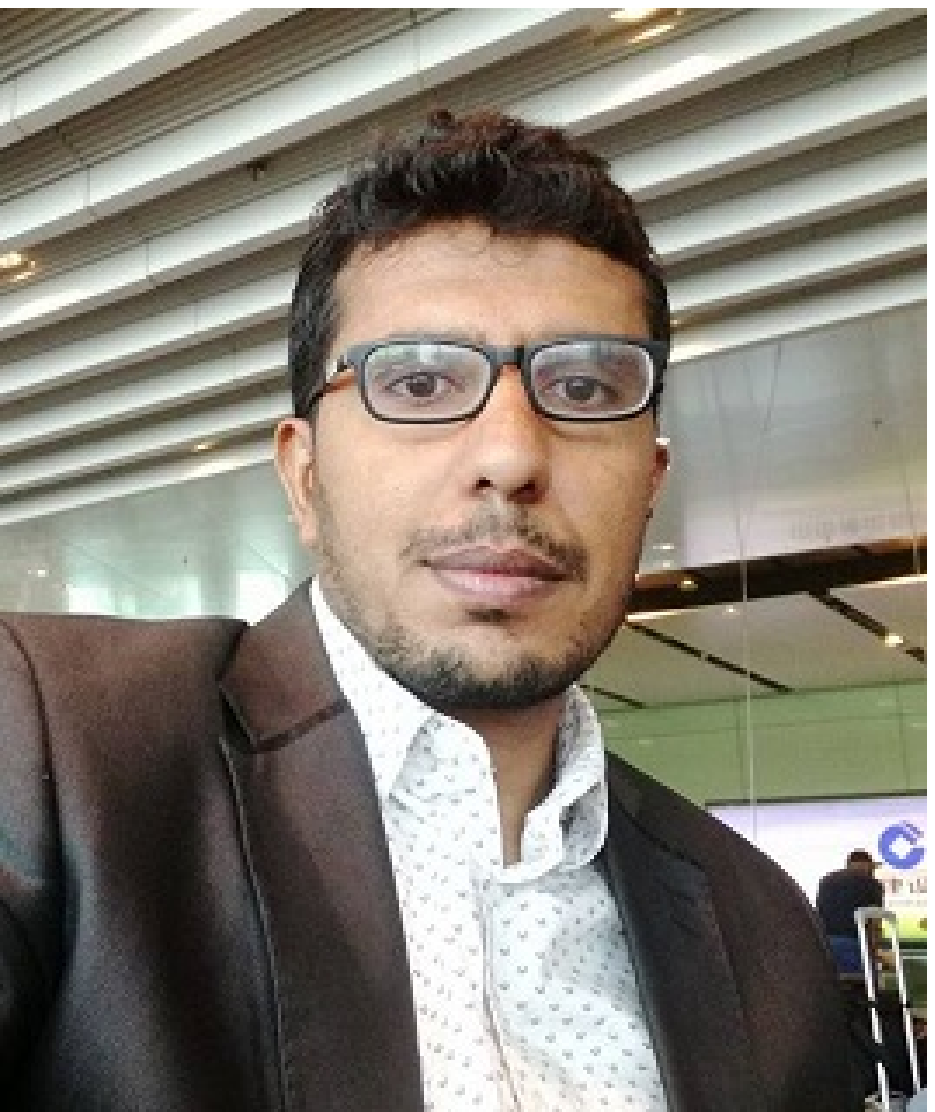}}]{Mohanad Obeed}
(S'17) received the B.Eng. degree in computer and communication engineering from Taiz University, Taiz, Yemen, in 2008, the M.Sc. degree in electrical engineering from  King Fahd University of Petroleum and Minerals
(KFUPM), Dhahran, Saudi Arabia, in 2016. 

He is currently pursuing the Ph.D. degree at KFUPM, Dhahran, Saudi Arabia. His research interests include visible light communications, cooperative communication, resource allocation, convex optimization, physical layer security, and energy harvesting.
\end{IEEEbiography}
\begin{IEEEbiography}[{\includegraphics[width=1in,height=1.25in,clip,keepaspectratio]{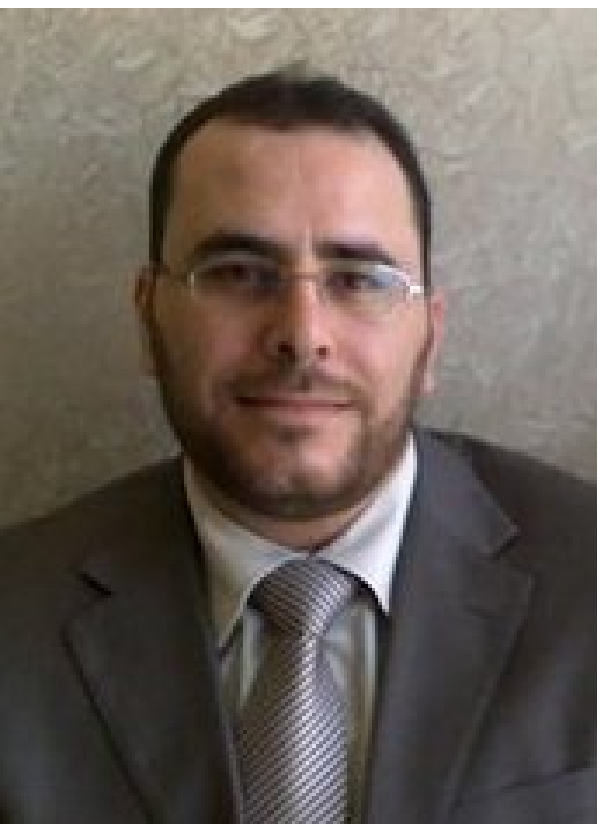}}]{Anas M. Salhab}
(S'11-M'14-SM'17) received the B.Sc. degree in electrical
engineering from Palestine Polytechnic University, Hebron, Palestine,
in 2004, the M.Sc. degree in electrical engineering from Jordan
University of Science and Technology, Irbid, Jordan, in 2007, and the
Ph.D. degree from King Fahd University of Petroleum and Minerals
(KFUPM), Dhahran, Saudi Arabia, in 2013. From 2013 to 2014, he
was a Postdoctoral Fellow with the Electrical Engineering Department,
KFUPM. He is currently an Assistant Professor and the Assistant
Director of the Science and Technology Unit with the Deanship of
Scientific Research, KFUPM. His research interest spans special topics
in modeling and performance analysis of wireless communication
systems, including cooperative relay networks, cognitive radio relay networks,
free space optical networks, visible light communications, and
co-channel interference. He was selected as an Exemplary Reviewer
by the IEEE WIRELESS COMMUNICATIONS LETTERS for his
reviewing service in 2014. Recently, he received the KFUPM Best
Research Project Award as a Co-investigator among the projects in 2013/2014 and 2014/2015.
\end{IEEEbiography}
\begin{IEEEbiography}[{\includegraphics[width=1in,height=1.25in,clip,keepaspectratio]{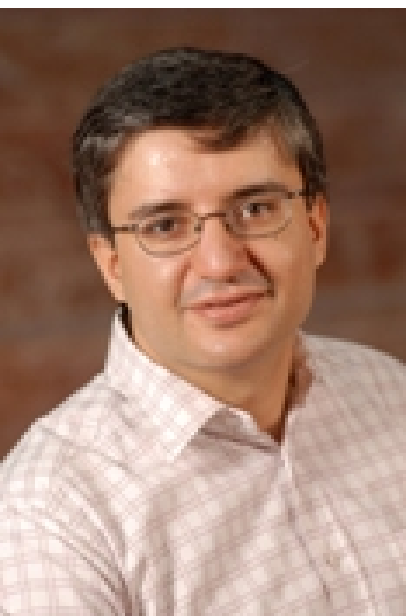}}]{Mohamed-Slim Alouini}
(S'94, M'98, SM'03, F'09) was born in Tunis, Tunisia. He received the Ph.D. degree in Electrical Engineering
from the California Institute of Technology (Caltech), Pasadena,
CA, USA, in 1998. He served as a faculty member in the University of Minnesota,
Minneapolis, MN, USA, then in the Texas A\&M University at Qatar,
Education City, Doha, Qatar before joining King Abdullah University of
Science and Technology (KAUST), Thuwal, Makkah Province, Saudi
Arabia as a Professor of Electrical Engineering in 2009. His current
research interests include the modeling, design, and
performance analysis of wireless communication systems.
\end{IEEEbiography}
\begin{IEEEbiography}[{\includegraphics[width=1in,height=1.25in,clip,keepaspectratio]{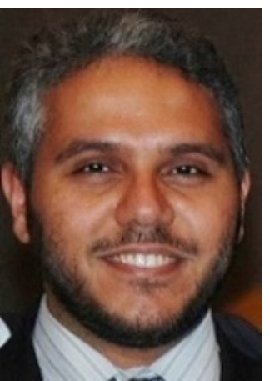}}]{Salam A. Zummo}
 (M'00-SM'08) received the B.Sc. and M.Sc. degrees in electrical engineering from the King Fahd University of Petroleum and Minerals (KFUPM), Dhahran, Saudi Arabia, in 1998 and 1999, respectively, and the Ph.D. degree from the University of Michigan, Ann Arbor, USA, in 2003. He is currently a Professor with the Electrical Engineering Department, KFUPM. Prof. Zummo has six issued U.S. patents and authored over 100 papers in reputable journals and conference proceedings. His research interests are in the area of wireless communications, including cooperative diversity, cognitive radio, multiuser diversity, scheduling, MIMO systems, error control coding, multihop networks, and interference modeling and analysis in wireless systems. He received the Saudi Ambassador Award for early Ph.D. completion in 2003, and the British Council/BAE Research Fellowship Awards in 2004 and 2006. He also received the KFUPM Excellence in Research Award from 2011 to 2012.
\end{IEEEbiography}

\end{document}